# Delivery of Dark Material to Vesta via Carbonaceous Chondritic Impacts

Vishnu Reddy
Max Planck Institute for Solar System Research, Katlenburg-Lindau, Germany
Department of Space Studies, University of North Dakota, Grand Forks, USA
Email: reddy@mps.mpg.de

Lucille Le Corre
Max Planck Institute for Solar System Research, Katlenburg-Lindau, Germany

David P. O'Brien
Planetary Science Institute, Tucson, Arizona, USA

Andreas Nathues
Max Planck Institute for Solar System Research, Katlenburg-Lindau, Germany

Edward A. Cloutis
Department of Geography, University of Winnipeg, Manitoba, Canada

Daniel D. Durda
Southwest Research Institute, Boulder, Colorado, USA

William F. Bottke
Southwest Research Institute, Boulder, Colorado, USA

Megha U. Bhatt
Max Planck Institute for Solar System Research, Katlenburg-Lindau, Germany

David Nesvorny
Southwest Research Institute, Boulder, Colorado, USA

Debra Buczkowski
Applied Physics Laboratory, Johns Hopkins University, Laurel, Maryland, USA

Jennifer E. C. Scully
Institute of Geophysics and Planetary Physics, University of California Los Angeles, Los Angeles
California, USA

Elizabeth M. Palmer
Institute of Geophysics and Planetary Physics, University of California Los Angeles, Los Angeles
California, USA

Holger Sierks
Max Planck Institute for Solar System Research, Katlenburg-Lindau, Germany

Paul J. Mann
Department of Geography, University of Winnipeg, Manitoba, Canada

Kris J. Becker
Astrogeology Science Center, U.S. Geological Survey, Flagstaff, Arizona, USA

Andrew W. Beck
Department of Mineral Sciences, Smithsonian National Museum of Natural History, 10th and Constitution NW, Washington, DC, USA

David Mittlefehldt
Astromaterials Research Office, NASA Johnson Space Center, Mail Code KR, Houston, Texas, USA

Jian-Yang Li
Department of Astronomy, University of Maryland, College Park, Maryland, USA

Robert Gaskell
Planetary Science Institute, Tucson, Arizona, USA

Christopher T. Russell
Institute of Geophysics and Planetary Physics, University of California Los Angeles, Los Angeles California, USA

Michael J. Gaffey
Department of Space Studies, University of North Dakota, Grand Forks, USA

Harry Y. McSween
Department of Earth and Planetary Sciences, University of Tennessee, Knoxville, Tennessee, USA

Thomas B. McCord
Bear Fight Institute, Winthrop, Washington, USA

Jean-Philippe Combe
Bear Fight Institute, Winthrop, Washington, USA

David Blewett
Applied Physics Laboratory, Johns Hopkins University, Laurel, Maryland, USA

Pages: 58
Figures: 15
Tables: 2

**Proposed Running Head:** Dawn at Vesta, Dark Material

**Editorial correspondence to:**
Vishnu Reddy
Max-Planck Institute for Solar System Research,
37191 Katlenburg-Lindau,
Germany.
 (808) 342-8932 (voice)
reddy@mps.mpg.de

**Abstract**

NASA's Dawn spacecraft observations of asteroid (4) Vesta reveal a surface with the highest albedo and color variation of any asteroid we have observed so far. Terrains rich in low albedo dark material (DM) have been identified using Dawn Framing Camera (FC) 0.75 μm filter images in several geologic settings: associated with impact craters (in the ejecta blanket material and/or on the crater walls and rims); as flow-like deposits or rays commonly associated with topographic highs; and as dark spots (likely secondary impacts) nearby impact craters. This DM could be a relic of ancient volcanic activity or exogenic in origin. We report that the majority of the spectra of DM are similar to carbonaceous chondrite meteorites mixed with materials indigenous to Vesta. Using high-resolution seven color images we compared DM color properties (albedo, band depth) with laboratory measurements of possible analog materials. Band depth and albedo of DM are identical to those of carbonaceous chondrite xenolith-rich howardite Mt. Pratt (PRA) 04401. Laboratory mixtures of Murchison CM2 carbonaceous chondrite and basaltic eucrite Millbillillie also show band depth and albedo affinity to DM. Modeling of carbonaceous chondrite abundance in DM (1-6 vol%) is consistent with howardite meteorites. We find no evidence for large-scale volcanism (exposed dikes/pyroclastic falls) as the source of DM. Our modeling efforts using impact crater scaling laws and numerical models of ejecta reaccretion suggest the delivery and emplacement of this DM on Vesta during the formation of the ~400 km Veneneia basin by a low-velocity (<2 km/sec) carbonaceous impactor. This discovery is important because it strengthens the long-held idea that primitive bodies are the source of carbon and probably volatiles in the early Solar System.

# 1. Introduction

NASA's Dawn spacecraft entered orbit around asteroid (4) Vesta in July 2011 for a year-long mapping project using its suite of three instruments. The Dawn Framing Cameras are a pair of identical instruments (one for cold redundancy) (Sierks et al. 2011) that have imaged the entire visible surface of the asteroid in clear (panchromatic) and seven colors (0.44-1.0 μm) with a resolution up to ~20 meter/pixel. During the approach phase, the entire surface was imaged in three successive rotational characterizations (RC1-3) at resolutions of 9.06 km/pixel, 3.38 km/pixel and 487 m/pixel, respectively. Following orbit insertion, Vesta was mapped from Survey, high-altitude mapping orbit (HAMO), and low altitude mapping orbit (LAMO) at resolutions of 250 m/pixel, 60 m/pixel, and 20 m/pixel, respectively.

The surface of Vesta as imaged by Dawn FC revealed a surface that is unlike any asteroid we have visited so far with a spacecraft. Albedo and color variations of Vesta are the most diverse among the objects in the asteroid belt, with a majority of these variations linked to distinct compositional units on the asteroid's surface (Reddy et al. 2012a). Whereas Hubble Space Telescope (Thomas et al. 1997, Binzel et al. 1997) and ground based (Gaffey, 1997) observations of Vesta had shown a large hemispherical scale dichotomy (Reddy et al. 2010), Dawn FC color data showed smaller regional and local scale albedo variations. Vesta's Western hemisphere on average has lower albedo than the Eastern hemisphere (Reddy et al 2012a). The large low albedo region extends from 90° to 200° longitude and 64° S to 16° N latitude and was originally interpreted as ancient howarditic terrain from ground-based spectral observations (Gaffey, 1997). All coordinate used in this paper conform to the Claudia system used by the Dawn science team. Bright material extending on either side of this dark region has been interpreted as impact ejecta from the formation of the Rheasilvia basin at the South Pole (Reddy et al. 2012a).

Preliminary analysis by Reddy et al. (2012a) showed carbonaceous chondrite contaminants as a possible source of the dark material. This line of evidence is supported by laboratory study of HED meteorites (e.g., Buchanan et al. 1993; Zolensky et al. 1996), and ground based spectral observations that suggested the presence of a 3-µm feature probably due to $OH^-$ in carbonaceous chondrites (Hasegawa et al. 2003; Rivkin et al. 2006). A preliminary geomorphological description of DM on Vesta is presented in Jaumann et al. (2012). More recently McCord et al. (2012) confirmed this hypothesis that dark material on Vesta is indeed related to carbonaceous chondritic clasts in HED meteorites. Here we presented a more detailed analysis testing this hypothesis and further constrain the origin, abundance and delivery mechanism for DM on Vesta. We also investigated alternate compositional affinities for DM.

**2. Description of Dark Material**

The Dawn Framing Camera (FC) discovered DM on Vesta (Reddy et al. 2012a). These enigmatic surface features in the form of low albedo units were first observed during RC3 at a resolution of ~487 m/pixel. Whereas they are ubiquitous in the low albedo hemisphere, their distribution is global with several occurring on the putative Rheasilvia ejecta units. Features associated with DM can be broadly classified into (A) those associated with impact craters (in the ejecta material and/or on the crater walls and rims); (B) flow-like deposits or streaks/rays commonly associated with topographic highs; and (C) dark spots. Figures 1-3 show examples of each type of feature using images obtained during HAMO and LAMO phases at a resolution of 20 to 60 m/pixel. Here we describe each group of features associated with DM to provide geological context for understanding their origin. We selected the 15-km diameter impact crater Cornelia (9.3S°, 225.5°E), which has a dark ejecta blanket and a prominent DM deposit on the crater walls and floor, as a test site for our analysis.

*2.1 Impact Craters*

A majority of DM on Vesta is found in and around impact craters. Figures 1 and 2 show the first of the two types of impact craters. In Fig. 1A, DM appears in the ejecta blanket left of the rim of the 58-km Marcia crater and also along the inner walls (see white arrows). It appears that the inner crater wall was likely covered by the DM seen on the ejecta blanket on the left of the crater rim, but subsequent downslope movements of fresher bright material has buried and mixed with the original DM. The exact cause for this mass wasting is unknown but seismic shaking triggered by nearby impacts could be a possible explanation. Dark material on the crater walls appears to have a 'tadpole' shape (see white arrows in Fig. 1B) due to bright material (likely more recent and less space weathered) flowing around them and covering part of their corresponding debris flow. Topographic features (i.e. pile of boulders or bright outcrops of brecciated material) along the crater walls seem to divert the bright granular material in channel-like features around the DM near the rim of the crater. Further down the slope, the bright debris flows spread out (indicated by dashed white arrows in Fig. 1B) forming lobate deposits (talus cones) that cover older dark deposits causing these 'tadpole' shaped features. Similar morphology can also be observed on the Moon: a classic example is the Diophantus crater on the western edge of Mare Imbrium that exhibits avalanche scars due to down slope movement of darker pyroclastic material. The DM 'tadpole' features are protected by the topography (in this case, outcrops of bright material) and in this way preserved from being completely covered by the subsequent debris flow. The layer forming the bright outcrops and the underlying dark debris flows produce this appearance of a lithology constituted by sub-surface deposit of DM with a boundary more or less contiguous or parallel to the crater rim. Characteristics of the bright layer might cause the variations in position and inclination of this apparent stratification clearly visible in the case of Marcia and partially visible in Numisia (Fig. 2). This layer might represent either an ancient DM lithology or the remnants of material deposited by the impact event and covered afterwards by debris flows from the crater wall.

Figure 2A shows another example of DM associated with an impact crater (Numisia). Unlike Marcia crater, some DM debris flows appear almost intact on the Northern part of the crater wall. Dark material is also visible in the ejecta blanket (see white arrows in Fig. 2A) and seems to have contributed to the DM falling along the crater wall by mass wasting. The intact DM debris flows (Fig. 2B) are originating from the northern rim of the crater and are flowing downslope around the topographic highs composed of bright material outcrops. In another part of the crater wall, smaller dark deposits are detectable and have the same 'tadpole' morphology that is found in Marcia crater (indicated by white arrows in Fig 1B). As seen for Marcia, these features are surrounded by bright and medium albedo debris flows coming from the rim crest (see dashed white arrows in Fig. 1C). These subsequent mass-wasting features have covered and mixed with the pre-existing DM deposits, forming lobes when they reached the crater floor. They are able to cover part of the bright outcrops as well and efficiently bury some parts of the crater wall.

Small impact craters on the eject blanket give us an indication of the thickness of the DM (Fig 2B). The interiors of some small craters seem to have the same brightness as the underlying ejecta blanket and some of them appear to have even darker albedo inside the craters and in their ejecta (see black arrows in Fig. 2A), exposing material that has not been buried by regolith or altered by space weathering. This suggests that either the DM, in this area, has a depth at least as deep as the small craters or the small craters were preexisting and then buried by the ejecta material. In contrast, other small craters of similar diameter located outside the DM ejecta have brighter interiors (see dashed white arrows in Fig. 2A). In addition, further away from the rim, a few small craters that formed on the dark ejecta blanket have bright ejecta indicating that the dark ejecta blanket is likely thinner with increasing distance from the crater rim (see dashed black arrows in Fig. 2A).

Some impact craters have ejecta that consists of DM mixed with target material in a subtle radial scour texture (Fig. 3A-B). Dark material in such ejecta typically forms thin radial rays extending asymmetrically from the crater floor. The asymmetry of the ejecta could be due to impact on a slope or an oblique impact. These craters also exhibit irregular concentrations of DM along the crater rim and walls. Other examples of DM, such as that at the Rubria crater (7.4S, 18.4E), have discontinuous ejecta blankets, usually as patches.

Other craters have dark interiors and ejecta (e.g. Fig. 3C). In these cases, DM is commonly distributed radially around the crater rim. A small impact crater (Fig. 3C) is located on the ejecta blanket of the 60-km diameter impact crater Marcia. Its dark ejecta suggest either that the impactor was a dark object or that a shallow subsurface low albedo layer was excavated.

*2.2 Dark Spots*

Dark spots appear as low albedo pits and can be found on the ejecta blankets of large impact craters such as the 30-km diameter Numisia crater (Fig. 2A). A majority of these dark spots are <1 km in diameter with some showing no evidence for a central craterlet. Fig. 3D shows examples of dark spots surrounding two larger impact craters. Dark spots are either the result of secondary impacts in cases where they have a discernable craterlet, or the result of ejecta material originating from nearby impact craters. Some dark spots appear independent of impact craters. An example is located in the region centered at 30°S, 120° E where dark spots are ubiquitous and unassociated with any larger impact craters.

*2.3 Topographic Highs*

Dark material deposits associated with topographic highs are found at two locations on Vesta: Aricia Tholus (Fig. 4) with a diameter of 37 km and Lucaria Tholus (Fig. 5) with a diameter of approximately 22 km. Aricia Tholus has DM in the form of rays (see white arrows in Figs. 4 A-B) radially emanating from a central crater that could represent an impact crater on a hill. In the case of Lucaria Tholus, located just south of the set of giant equatorial grooves, DM is draping part of a short ridge. Dark material forks out from this ridge crest into the floor of nearby grooves. The downslope movement has formed a long flow-like feature to the east (indicated by white arrows in Figs. 5A-D).

Similarly, a debris flow has been deposited to the south of this edifice, forming another branch of DM in the downslope direction (indicated by black arrows in Figs. 5A-D).

## 3. Quantifying Color Parameters of Dark Material

*3.1 Data Processing*

Framing camera images are processed in three levels (1a-1c) in which the data stream received on the ground (level 0) is converted to PDS format images (level 1a) that contain unprocessed, uncalibrated digital values from 0 to 16383 DN. Level 1a data are converted to level 1b as radiometrically calibrated PDS-compliant format images. The level 1b data are converted to reflectance (I/F) by dividing the observed radiance by solar irradiance from a normally solar-illuminated Lambertian disk. Dawn FC images are affected by infield stray light that is corrected on the level 1b I/F data. The stray light reduction algorithm removes this effect in the color data where the in-field stray light can be up to 15% depending on the filter. The stray-light corrected data is then input to the ISIS color pipeline. ISIS (Integrated Software for Imagers and Spectrometers) (Anderson et al. 2004) is a UNIX-based program developed and maintained by USGS. ISIS performs photometric correction of the color data to standard viewing geometry (30° incidence and 0° emission angles) using Hapke functions derived from disk-integrated ground-based telescopic observations of Vesta and Vestoids and spacecraft data acquired during the approach phase (RC1-3 and Survey). The photometrically corrected data is map projected and coregistered to align the seven color frames in order to create color cubes prior to analysis. A detailed description of the FC data processing pipeline is presented in Reddy et al. (2012a).

*3.2 Color Parameters*

The albedo of DM is a key factor in its identification and displays the highest visual contrast in the 0.75 μm filter. The albedo trend between various surface color units on Vesta appears to form a continuum, with the DM having the lowest values (0.09-0.15). A typical spectrum of Vesta is dominated by the mineral pyroxene with characteristic absorption bands at 0.9 and 2.0 μm. Pyroxene on the surface of Vesta is a product of igneous activity. Four of the seven Dawn FC color filters (0.44-1.0 μm) cover the 0.9-μm pyroxene band. The intensity of the 0.9-μm pyroxene absorption is also affected by the presence of DM (Le Corre et al. 2011; Reddy et al. 2012a). We used *R(0.75)/R(0.92)*, where *R* is the reflectance at wavelength *λ*(μm), as a proxy for quantifying the band depth, which is usually an indicator of iron content, particle size, space weathering, and abundance of opaques like carbon/metal (Reddy et al. 2012).

The ratio of 0.98/0.92-μm is called the Eucrite-Diogenite (ED) ratio as it helps to distinguish the two meteorite types. Eucrites have higher iron and calcium contents in pyroxenes than diogenites. Hence, the 0.90-μm pyroxene band is shifted toward longer wavelength for eucrites compared to diogenites. This leads to a 0.98/0.92-μm ratio that is ≤1 for eucrites, but >1 for diogenites. Given the collisional mixing of the compositional units on Vesta, the ED ratio at best helps identify eucrite-rich or diogenite-rich polymict terrains. By using this ED ratio one can determine if the target materials of the impact craters with DM are rich in eucrite or diogenite.

*3.3 Albedo: How dark is dark material?*

The albedo of the DM is a key factor in its identification along with surface temperature. Reflectance of a surface varies with incidence, emission and phase angles and therefore photometric correction of the data is required prior to any scientific analysis.

In order to quantify the albedo of DM, it is also important to quantify the albedo of the brightest material on Vesta. Reddy et al. (2012a) noted that the brightest material (in clear filter) has a geometric albedo of ~0.67 whereas the darkest material has a geometric albedo of only ~0.10. This albedo range is the highest among asteroids that have been observed from the ground or visited by spacecraft.

To quantify the albedo change due to the presence of the DM, we have chosen the 0.75-μm filter as it is one of the least affected by infield stray light. This filter also provides a good contrast between the DM and the other units. Figure 6A shows an image of two impact craters with DM ejecta near Divalia Fossa (Rubria and Occia craters). A profile drawn across (red line) the ejecta of the crater to the lower right is plotted in Fig. 6B. The albedo at 0.75 μm of the gray ejecta adjacent to the DM is ~0.22 but drops steeply to ~0.12 over the DM ejecta (~45 % decrease in reflectance). As a comparison, the brightest material on Vesta has an albedo of ~0.45 for the same filter. Figs. 6A-B provide evidence that the albedo of the DM at 0.750 μm is significantly lower than its immediate surroundings and is an important parameter for identification.

*3.4 Band Depth: How deep is the 0.9- μm pyroxene band?*

Reddy et al. (2012a) showed that the average spectrum for DM has a lower band depth (23%) compared to Vesta's global average (46%). Figures 7A-C show color-coded maps of various color parameters for the 15-km diameter impact crater Cornelia with DM. Figure 7A shows the image of the impact crater in 0.75 μm filter and Fig. 7B shows the ratio of 0.75/0.92 μm filters (proxy for band depth). The color code is red for areas with deepest 0.9-μm pyroxene bands and blue for areas with weakest bands. Figure 7B confirms the spectral properties that we described earlier from the color spectra: the DM suppresses the 0.9-μm absorption band and the intensity of the suppression could be a function of DM in the surface mixture.

*3.5 Eucrite or Diogenite*

Identification of the target material with which the DM is mixed can help decipher its origin. As noted earlier, ED ratio helps to identify eucrite-rich or diogenite-rich polymict terrains. In the ED color-coded image of Cornelia (Figure 7C), yellow to red areas are those richer in diogenite-like material. Dark material seen in Fig. 7A corresponds to blue in the ED color-coded map suggesting the target material is a mixture of eucrite (ED ratio ~1) and diogenite (ED ratio >1), but with more diogenite. However, it must be noted that the presence of DM might affect the ED ratio leading to false identification of eucrite-rich terrains.

*3.6 Summary of Color parameters*

Based on the results inferred from Figs. 6A-B and 7A-C, it is clear that the DM has lower albedo (~0.08-0.15), and shallower 0.9-μm band depth (~23%) compared to the average spectrum of Vesta. We have performed similar analysis on other DM terrains (i.e. for dark spots and for DM deposited on topographic highs) and the observed trends in color parameters are identical. This suggests that a majority of the DM we see on Vesta shares common properties (i.e. composition, grain size) and origin. In an effort to constrain the origin of this DM, we have created a matrix that describes the color properties (albedo, band depth) of all the possible analog materials we identified (Table 1). This matrix lists processes or analogs that have been observed either in the laboratory or on other planetary bodies. By solving for this matrix we hope to narrow down the best source for the DM.

**4. Endogenous Sources of Dark Material**

*4.1 Pyroclasts from Vesta*

The paucity of pyroclastic products such as glass and opaques in HEDs is well established (Wilson and Keil, 1997; Keil, 2002). Magma gas content of HED meteorites (Mittlefehldt 1987; Grady et al. 1996, 1997) is far lower (tens of parts per million) than what would be required (>3.8 wt.%) for eruption speed to exceed the escape velocity (~390 m s-1) of Vesta (Wilson and Keil, 1991). However, this low gas content is sufficient to allow explosive activity to create pyroclastic products.

Wilson and Keil (1997) explained the paucity of pyroclastic products in HED by modeling eruption speeds and trajectories of pyroclastic magma droplets. They suggested that pyroclastic magma droplets formed in lava fountains are extremely optically dense and hence will undergo limited radiative cooling during their trajectory to the surface. As a consequence, the droplets would merge to form lava ponds rather than deposition of glass spherules like those found on the Moon, thereby explaining the lack of traditional pyroclastic products in HEDs. Wilson and Keil (1997) estimated that <1% of the erupted lava would form pyroclastic fall deposits similar to glass beads. Moreover, they also predicted that minerals crystallizing from these lava ponds would be indistinguishable from those formed by extrusive lava flows (i.e. basaltic eucrites). Therefore any lava flows erupting from a fire fountain would be spectrally identical to basaltic eucrites.

Aside from the low abundance of vestan pyroclastic fall deposits, the identification of pyroclasts in the HEDs is also challenging (Singerling and McSween, 2012). Compared to lunar pyroclasts, which exhibit distinct increase in magnesium and decrease in aluminum along with a typical glassy texture, vestan pyroclasts with similar properties can easily be mistaken for impact melts. Comparison with lunar pyroclasts shows that the weakest band depth found in lunar pyroclasts (i.e., Apollo 15 green glass) is nevertheless still stronger than the band depth of DM.

*4.2 Eucrites with Opaques*

The presence of opaque phases in eucrites (chromite, ilmenite, troilite, and metal) (Mayne et al. 2010) reduces both albedo and band depth (Miyamoto et al. 1982). The intensity of these effects is a function of the grain size and modal abundance of the opaque phases. A band depth vs. albedo plot (Fig. 8) of three opaque-rich eucrites (MET 01081, BTN 00300, Chervony Kut) (Mayne et al. 2010) and vestan DM reveals that all three have higher albedo and band depth compared to the DM. Due to this mismatch, these meteorites are unlikely to be DM analogs. Furthermore, the abundance of opaque-rich meteorites among unbrecciated eucrites provides additional support for this conclusion.

Opaque-rich eucrites are rare (only 3 of 29) (Mayne et al. 2010) and the abundance of opaques (i.e., chromite/ulvospinel, ilmenite, sulfides, and FeNi metal) in these three eucrites is ≤2% (typically ~1%). Thus, opaques are rare among the unbrecciated eucrites and if present, are only observed in trace amounts even for opaque-rich eucrites. At higher abundances, chromite and ilmenite would exhibit absorption features in the 0.5-0.6 μm region (Adams, 1974; Cloutis et al. 2004), which are not observed in FC spectra of the DMs. Abundant troilite or metal (Britt and Pieters 1989) would lead to progressively redder slopes, again a characteristic that is not observed in DM FC spectra (Cloutis et al. 2009). Apart from the three opaque-rich eucrites, ALHA81001 is a quenched basaltic eucrite that has <1% fine-grained ilmenite scattered throughout the meteorite. This meteorite's lower spectral contrast has been attributed to its fast cooling rate rather than the presence of ilmenite (Mayne et al. 2010). Our independent study (Reddy et al. 2012b) to detect opaque-rich terrains similar to ALHA81001 using the chromium absorption band in pyroxene has been fruitless, suggesting that it is not a dominant assemblage.

*4.3 Impact Shock and Impact Melt*

Impact shock can lower the albedo and reduce the band depth of the original target material (Gaffey, 1976; Adams et al. 1979). In shock-blackened ordinary chondrites, the dispersion of micron-scale nickel iron metal and troilite is the primary cause of these effects (Britt and Pieters, 1989). The metal and silicate components of black ordinary chondrites are identical to their unshocked counterparts. However, black chondrites have albedos typically <0.15 compared to 0.15-0.30 for intact ordinary chondrites. $^{40}$Ar-$^{39}$Ar measurements of eucrites suggest at least four large impact events occurred between 4.1 and 3.4 Ga that were capable of resetting Ar ages (Bogard and Garison, 2003). Those impacts would be strong enough to significantly shock the target material, as evidenced in HED meteorites, and could potentially produce shock-blackened eucrites like the observed shock-blackened ordinary chondrites. However, very few shock-blackened eucrites are known in the terrestrial meteorite collections.

Another product of the impact process is glass. Although a few howardites have impact melt glass between 15-20 vol% (Desnoyers and Jerome, 1977), in general, glass represents <1 vol% of known HED meteorites. An impact melt clast from eucritic breccia Macibini contains 50% devitrified glass and 50% silicates (Buchanan et al. 2000) and the Padvarninkai meteorite is the most heavily shocked eucrite known with most of the plagioclase converted to maskelynite with significant impact melt glass (Hiroi et al. 1995). LEW85303 is a polymict eucrite with an impact melt matrix. Jiddat al Harasis (JaH) 626 is a polymict eucrite that experienced significant impact shock and melt quenching (Irving, 2012). To test the affinity of DM to shocked eucrites and impact melt we obtained spectra of the Macibini clast and the Padvarninkai eucrite from RELAB database. We also measured the visible-near-IR spectra of JaH 626. All spectral data were resampled to Dawn FC filter band passes to obtain their color spectra.

The albedo of a <45 μm powder of JaH 626 at 0.75 μm is comparable to other polymict eucrites (albedo of 0.32) (Fig. 9). The albedo of JaH 626 (and other eucrites) drops significantly with increasing particle size with the lowest albedo being 0.12 at 250-500 μm. Band depth also decreases with increasing grain size suggesting that shocked impact melt-rich material could be a possible analog for some DM units. However, albedo and band parameters of JaH 626 appear higher than those of DM. The mismatch in spectral band parameters coupled with the paucity of shock-blackened eucrites in meteorite collections makes it unlikely for JaH 626-type material to be a possible analog for the DM on Vesta.

The albedo of the Pavarninkai impact melt sample with 25-45 micron grain size (albedo is 0.15) is comparable to the DM observed on in Cornelia crater. However, the band depths of both grain sizes are lower and they plot below the DM trend in the band depth vs. albedo plot (Fig. 9). Macibini impact melt clast and LEW85303 have albedo (0.24, 0.36) and band depth that is significantly higher than DM. Like JaH 626 the color parameters of impact melts are significantly different from DM observed on Vesta making them unlikely analogs.

## 5. Exogenous Sources of Dark Material

*5.1 Carbonaceous Chondrite Impactor*

Exogenous sources (e.g., a carbonaceous chondrite impactor) could explain the DM seen on Vesta. The presence of exogenous carbonaceous chondrite meteorite clasts in HED meteorites from Vesta is well documented (e.g., Buchanan et al. 1993; Zolensky et al. 1996). Unshocked CM-like carbonaceous chondrite clasts have also been identified in other meteorite types, including H chondrite meteorite breccias, where they make up the most abundant variety of foreign materials in those rocks (e.g., Rubin and Bottke 2009). Carbonaceous material can also occur as microclasts (25-800 μm) similar to micrometeorites on Earth (Gounelle et al. 2003).

In HEDs, carbonaceous chondrite clasts generally comprise up to 5 vol. % (Zolensky et al. 1996), but on rare occasions can be 60 vol. % of howardites (Herrin et al. 2011). Antarctic howardites Mt. Pratt (PRA) 04401 and 04402 along with Scott Glacier (SCO) 06040 have 10-60% carbonaceous chondrite clasts (Herrin et al. 2011). A vast majority (83%) of these are CM2 carbonaceous chondrites followed by CR2 (15%). Both of these meteorite types are aqueously altered (Zolensky et al. 1996). Clasts are primarily composed of a phyllosilicate matrix (mostly serpentine) with Fe-Ni sulfides, carbonates, hydrated sulfates, organics, and $Fe^{3+}$-bearing phases, and inclusions of olivine, pyroxene, relict chondrules and calcium-aluminum-rich inclusions. Although some of these clasts have been heated and dehydrated during impact (~400°C), a majority are still hydrated (containing $H_2O$- and/or OH-bearing phases) (Zolensky et al. 1996).

In order to test the possibility of carbonaceous chondrite impactors being the source of DM on Vesta, we studied color properties of eucrite (Millbillillie) and CM2 carbonaceous chondrite (Murchison) mixtures (Le Corre et al. 2011) and Antarctic howardite PRA04401. Figure 10A shows a spectrum of DM overlaid on color spectra of CM2+eucrite mixtures. Both data sets are normalized to unit at 0.75 μm. Band depth vs. albedo of DM along with those of PRA04401 and CM2+eucrite mixtures are plotted in Fig. 10B. Band depth and albedo values for CM2+eucrite mixtures (particle size <45 μm) are offset from the DM trend. This offset between the two could be a compositional effect as the CM2+eucrite mixture does not have a diogenitic component like PRA04401 howardite.

Adding a diogenitic component would likely increase the overall band depth of the mixture as diogenites typically have deeper pyroxene bands than eucrites. This would shift the overall CM2+eucrite mixture trend upward matching that of DM on Vesta. However, the band depth and albedo trends of PRA04401 and CM2+eucrite mixtures are the best possible matches for DM among all the analogs we have investigated so far. Areal and intimate mixture band depth-albedo trends run parallel on either side of the DM trend, suggesting that both processes are prevalent on Vesta's surface, with the former being similar to carbonaceous chondrite clasts (Zolensky et al. 1996) and the latter, microclasts (Gounelle et al. 2003). The same color parameters for PRA04401 (Fig. 10B) plot in the middle of the DM trend, confirming this interpretation.

*5.2. Abundance of CM2 in Dark Material*

Estimating the abundance of CM2 on the surface of Vesta is challenging because other types of surface processing and materials can mimic the color characteristics of DM. Although the effects of these secondary components (space weathering, impact melt, eucrites with opaques, shocked eucrites, iron abundance) on band depth and albedo are limited compared to carbonaceous chondrites, they nevertheless introduce uncertainties. Abundance was initially estimated using two equations based on laboratory calibration of CM2 and eucrite mixtures (Le Corre et al. 2011). We independently developed a third equation based on *R(0.75)/R(0.95)*. All three equations are based on intimate mixing model and their $R^2$ (regression) values are ~0.99.

$$CM2\ vol\% = 31531*\alpha^4 - 34378*\alpha^3 + 13655*\alpha^2 - 2469.3*\alpha + 192.48 \quad (Eq.\ 1)$$
$$\alpha = 0.75\ \mu m\ reflectance$$

$$CM2\ vol\% = -180.88*\beta^3 + 924.72*\beta^2 - 1604.5*\beta + 946.35 \quad (Eq.\ 2)$$
$$\beta = R(0.75)/R(0.92)$$

$$CM2\ vol\% = -178.32*\gamma^3 + 905.99*\gamma^2 - 1563.5*\gamma + 918.23 \quad (Eq.\ 3)$$
$$\gamma = R(0.75)/R(0.95)$$

Using these equations, we estimated the CM2 abundance range for several sites on Vesta with DM. The method we employed included the extraction of color spectra over the region of interest showing DM. Average, minimum (darkest) and maximum (brightest) spectra were calculated for each site and these values were used to obtain $\alpha$, $\beta$, and $\gamma$ parameters. CM2 abundance values were calculated for these nine parameters, and two average CM2 abundances were calculated. The first used all three parameters ($\alpha$, $\beta$, and $\gamma$) and the second used just $\beta$ and $\gamma$.

Determination of the exact abundance of carbonaceous material on the surface depends on the mixture model used (areal vs. intimate). We estimate abundance of carbonaceous chondrite material for the average surface of Vesta to be between <6 vol%, consistent with HED meteorites (Zolensky et al. 1996) and GRaND observations (Prettyman et al. 2012). Concentrations as high as ~50 vol% are noted in DM-rich areas which also agrees with 10-60 vol.% observed in howardites PRA 04401, PRA 04402 and SCO 06040 (Herrin et al. 2011). These localized concentrations are small and cover only an area of a few 10s of meters. These localized concentrations could be a result of non-uniform deposition/mixing/excavation of the impactor material and ejecta.

Phyllosilicates in CM2 carbonaceous chondrites contain adsorbed water as well as structural OH (Beck et al. 2010). Laboratory analysis of PRA04401 indicates that it has approximately 6 wt.% water, which produces an absorption feature at ~3-μm. If the DM found on Vesta is equivalent to PRA04401, then we can expect this spectral feature. Detection of such a feature by Dawn's VIR spectrometer associated with some DM strengthens this link (McCord et al. 2012). The independent mapping of hydrogen distribution provides further evidence by Dawn's GRaND instrument that show values consistent with hydrogen abundances in howardites containing CM2 carbonaceous clasts (Prettyman et al. 2012). The GRaND data shows clear correlation between H abundance and DM (Prettyman et al. 2012).

**6. Delivery of Dark Material**

The global distribution of the DM offers some insights into its origin, though there are a number of ways that material could have reached Vesta. Below we present several possible delivery mechanisms, and discuss some possible pros and cons to each one.

*6.1. Delivery of DM by a large carbonaceous chondrite impactor*

Evidence from HED meteorites suggests that at least two impacts (one each for CM2 and CR2 clasts) are necessary for the delivery of carbonaceous material. By mapping the location of morphological features associated with DM, we found a preferential concentration near the rim and on the floor of the ancient 450 km diameter Veneneia (52°S, 170°E) impact basin (Figs. 11-12) and a large low-albedo unit immediately to the north. In the sequence of events affecting the vestan south pole, we hypothesize that a low velocity impactor related to the Veneneia impact basin formation delivered some DM to Vesta. Presumably, this impactor would have been the one that delivered the CM2-type material, as it constitutes ~83% of all macro clasts in howardites[1]. The subsequent Rheasilvia impact (75°S, 301°E) erased about half of the Veneneia basin's structure and covered the other half with bright ejecta mantling (Fig. 12). In this scenario, more recent impacts excavated the sandwiched layers of bright Rheasilvia ejecta material, DM from Veneneia impactor, and some in situ regolith. These layered deposits are visible on crater walls (Fig. 1A and 2A), and some smaller impact craters also expose a possible sub-surface regional dark deposit layer (Fig. 3C). They are less consistent with DM excavated from craters on the floor of the Veneneia crater.

*6.1.1 Impact Velocity*

The impact velocity necessary for significant amounts of dark material to be emplaced on the surface of Vesta varies depending on the model parameters. Numerical studies indicate that Vesta's impact velocity distribution has a low-velocity tail where ~6% of impacts can occur at less than 2 km/s (Figs. 13A-B, Table 2). As discussed further in the following section, a low velocity impact would be advantageous for several reasons: Less impactor material would be lost to space in a low-velocity impact than in a high-velocity impact; and since lower velocity impacts form smaller craters, the remaining projectile material would be diluted in a smaller volume of ejecta and hence the ejecta would be richer in impactor material. It is also possible that during such an impact, significant quantities (cm size) of hydrated, unshocked CM2 material are spalled off the projectile (Rubin and Bottke, 2009) and achieved low relative velocity to land on Vesta (Bland et al. 2008). Dynamical simulations (Jutzi & Asphaug, 2011) show that the fraction of impactor material that can survive is strongly dependent on impact angle with ~90% for a head-on impact and ~50% for 45° impact angle. The fact that most CM2 clasts in howardites are still hydrated and unshocked suggests that they were deposited on Vesta at relatively low velocities (Zolensky et al. 1996; Buchanan et al. 1993; Gounelle et al. 2003). The VIR and GRaND data also support extensive hydration of DM relative to Vesta (Prettyman et al. 2012; McCord et al. 2012). A few CM clasts are shock heated and dehydrated during impact and probably accreted at higher impact velocities (Rubin and Bottke 2009) possibly producing pitted terrains observed in Marcia and Cornelia (Denevi et al. 2012).

On the other hand, this scenario makes use of several low probability events. The velocity threshold where a projectile leaves behind a substantial fraction of projectile material is unknown. Our calculations indicate that $V < 2$ km/s impacts should only occur once out of every 17 events, yet the Veneneia event also managed to make one of the largest craters observed on Vesta. If the $V < 2$ km/s velocity threshold were even lower, the odds would be significantly worse.

Lower velocities also mean that a larger projectile is needed to make a same-sized crater, at least when compared to mean values. Using crater-scaling relationships from Melosh (1989), we find that the ~400 km diameter Veneneia can be made by a roughly 30 km diameter projectile at 4.75 km/s, the mean impact velocity of main belt asteroids hitting Vesta, or a 50 km diameter projectile hitting at 2 km/s. Using the main belt size frequency distribution described in Bottke et al. (2005), we find that 30 km diameter impactors would hit Vesta about twice as often as 50 km diameter bodies.

Assuming the main belt size distribution has not substantially changed over solar system history, the mean interval between ≥30 km diameter impactors striking Vesta is ~4 Gy, whereas that for a 50 km projectile is ~8 Gy. These values are consistent with Veneneia-sized craters being rare on Vesta. Additional consideration must also be given to the fact that only some of the D > 30 km bodies in the main belt are carbonaceous chondrite-like.

Taken together, these calculations imply that either the Veneneia impact was an exceptional event, the velocity threshold needed to leave behind substantial quantities of unshocked projectile material is actually higher than 2 km/s, or that large impacts are not the source of the projectile material. From this point on, we will assume that one or both of the first two options are valid.

*6.1.2 Estimating Impactor Volume Fraction in Crater Ejecta*

Crater scaling laws can be used to estimate the fraction of impactor material that may survive in the ejecta from a crater. For crater volume as a function of projectile mass and impact velocity, Holsapple (1993, Table I) gives expressions for crater volume (m³) in a range of materials, including hard rock

$$V_s = 0.005 \, m \, v_{imp}^{1.65}$$
$$V_g = 0.48 \, m^{0.783} G^{-0.65} v_{imp}^{1.3}$$

soft rock

$$V_s = 0.009 \, m \, v_{imp}^{1.65}$$
$$V_g = 0.48 \, m^{0.783} G^{-0.65} v_{imp}^{1.3}$$

dry soil

$$V_s = 0.04 \, m \, v_{imp}^{1.23}$$
$$V_g = 0.14 \, m^{0.83} G^{-0.51} v_{imp}^{1.02}$$

and sand

$$V_g = 0.14\ m^{0.83}\ G^{-0.51}\ v_{imp}^{1.02}$$

where $V_s$ and $V_g$ are the crater volumes in the strength and gravity regimes, $m$ is the projectile mass in kilograms, $v_{imp}$ is the impact velocity in km/s, and $G$ is the surface gravity in units of Earth $g$ (for Vesta, $G$ = (0.22 m/s²)/(9.81 m/s²) = 0.22). Vertical impacts are assumed in these expressions, and the more likely case of non-vertical impacts (the average impact angle is 45 degrees) will result in somewhat smaller crater volumes. The final crater volume depends on whether strength or gravity limits the final size of the crater, and thus will be the smaller of $V_s$ and $V_g$.

As a first-order assumption of the ejecta volume, we assume that it is equal to the volume of the crater. This is likely an overestimate, as some of the crater volume is due to displacement rather than excavation, and some of the ejecta is likely lost to space rather than retained on the surface, but this upper-limit estimate will yield a lower-limit value of the volume fraction of projectile material that would be mixed with the crater ejecta in the following calculations.

For an impactor diameter D, the projectile mass and volume are

$$V_{imp} = \frac{1}{6}\pi D^3$$

$$m = \rho_{imp} V_{imp}$$

where we assume $\rho_{imp}$ = 2000 kg/m³ for carbonaceous impactors.

For the mass fraction of the projectile that escapes in sub-10 km/s impacts, Svetsov (2011, Eq. 8) finds

$$\frac{m_{esc}}{m} = 1 - (0.14 + 0.003\ v_{imp}) \ln v_{esc} - 0.9\ v_{imp}^{-0.24}$$

where $v_{esc}$ is 0.35 km/s for Vesta. This yields a retained mass fraction of projectile material that ranges from ~0.75 for 1 km/s impacts to ~0.5 for 5 km/s impacts. As a conservative estimate, we use a retained mass fraction of 0.5 for all calculations.

If we assume that the retained projectile material is uniformly mixed in with the retained ejecta material, then for a given cratering event we can estimate the volume fraction of projectile material (CC for carbonaceous chondrite) in the ejecta blanket as

$$f_{CC} = \frac{0.5\ V_{imp}}{\min(V_s, V_g)}$$

Figure 14 shows the carbonaceous chondrite volume fraction in the ejecta for a range of target materials, impactor diameters and velocities. Although there is some uncertainty about the best scaling law to use for Vesta, it will certainly lie within the range of materials used here. These are likely conservative estimates for several reasons. As noted above, we likely overestimate ejecta volume and use a low-end value of 0.5 for the fraction of impactor material retained on the surface, both of which will serve to give a lower value of $f_{CC}$. In addition, the crater volume calculations assume vertical impacts, and the more likely oblique impacts will eject less material and thus lead to a larger fraction of impactor material in the ejecta.

*6.1.3. Dynamical Modeling of Veneneia Ejecta*

To begin to test the hypothesis that DM on Vesta is distributed where ejecta from the Veneneia basin-forming event would be deposited on the asteroid's surface, we performed a simple dynamical simulation using the model described in Durda (2004), modified for Vesta conditions. Vesta was modeled as a triaxial ellipsoid with major axes of 578×560×458 km and a rotation rate of 5.342 hours. Ejecta from Veneneia was modeled as 10000 point masses launched from within the footprint of a 450-km-diameter crater at latitude -52 degrees and a model longitude with the appropriate relative position of the crater with respect to the asteroid's major axis. Ejecta speeds were distributed between 100-400 m/s, with the number of ejecta particles traveling at greater than a given speed determined by a power law. As the aim here was merely to examine in a qualitative sense the overall distribution of Veneneia ejecta across Vesta's surface under the combined effects of the asteroid's shape and rotation, we did not attempt to incorporate more detailed crater ejecta scaling laws at this stage.

Figure 15 shows the results of the dynamical model of the distribution of DM. Modeled ejecta are distributed generally south of the equator and to the west of the crater. The effects of rotation lead to a sharply delineated ejecta blanket boundary to the north of the crater and in the direction of the asteroid's spin, caused by the asteroid rotating 'out from under' the reaccreting ejecta. This boundary qualitatively resembles a sharp-edged boundary in the distribution of DM on Vesta located near latitude 0°, longitude 120°W. The modeled distribution does not produce reaccreted ejecta within the floor of the crater, but the dynamical model used here is an idealized representation of the crater ejecta process and was not intended to model the details of impact processesapplicable to basin-scale impacts. Still, the qualitative match between our dynamically modeled Veneneia ejecta distribution and the observed distribution of DM on Vesta suggests that the hypothesis of DM as Veneneia ejecta warrants further examination.

*6.1.4 Summary: Model vs. Observations*

CM2 material abundance estimated for several DM sites on Vesta show a range whose limits are consistent with our modeling results (Fig. 14). For the large low albedo unit, the average lower limit of CM2 abundance ranges from 8-11 vol. % consistent with a $f_{CC}$ range produced by an impactor arriving at 2 km/sec (Fig. 14). Localized concentrations as high as 53 vol. % are noted. Moderate impact craters like the 15-km diameter Cornelia and 30-km diameter Numisia have CM2 abundance range of 9-32 vol. % and 11-35 vol. %, respectively. Smaller impact craters like the one shown in Fig. 3C have localized CM2 abundances between ~12-33 vol. %. The lower end of these ranges for impactor material abundance are consistent with $f_{CC}$ estimated using scaling laws for velocities between 1-2 km/sec (Fig. 14).

Our estimates of CM2 abundance and distribution (Fig. 14) are on the conservative side, and the actual fractions are likely to be higher than our estimates. As noted in section 6.1.2, the ejecta volume is not equal to the crater volume as the material is only ejected from about ~1/3-1/2 of the depth of the transient crater. The rest of the crater volume is due to displacement of material (i.e. material getting pushed down by the impactor). This alone could increase the $f_{CC}$ in the ejecta by a factor of two or more compared to our estimates. A second factor is the distribution of the impactor material in the ejecta, which is likely to be non-uniform. This would imply that some parts of the surface within range of the Veneneia ejecta may contain little DM, whereas others may include larger concentrations. Both these factors, as well as others discussed in Sec. 6.1.2, suggest that very low velocity (<2 km/s) impacts may not be required, and that somewhat higher velocity (and hence higher probability) impacts could potentially give a large enough $f_{CC}$ to match the observations.

*6.2 Delivery of dark material by micrometeorites*

Another possible source of some DM to Vesta is micrometeorites. This DM layer would be more akin to a background global layer due to mixing with in-situ howardite regolith. Consider that most of the micrometeorites reaching Earth today are CI, CM, or CR-like bodies, with much of their mass being in the form of 100-200 micron diameter particles (Love and Brownlee 1995, Taylor et al. 2000, Love and Allton 2006). Numerical modeling work of the Zodiacal cloud indicates the vast majority of this material is produced by the disruption of Jupiter-family comets, or JFCs, whose ice-free component is likely to be similar to primitive carbonaceous chondrites (Nesvorny et al. 2009). Those tiny bodies that escape ejection via a close encounter with Jupiter evolve inward toward the Sun by Poynting-Robertson drag. As they cross the asteroid belt, many should have an opportunity to hit Vesta and other main belt asteroids.

This process has probably been going on in a similar fashion since the formation of the trans-Neptunian scattered disk about 4 Gy ago. We know that about 2% of lunar maria soils are made up of extralunar components that are most consistent with CI/CM-like chondrites (see Wasson et al. 1975). The current flux of micrometeorites hitting the Moon, 1600 tons per year, is sufficient to produce this fraction, provided it has been hitting the Moon at the same rate, more or less, over the last few billion years or so. Accordingly, we predict that Vesta should have also seen a steady flux of primitive micrometeorites over the same time interval.

Using the Nesvorny et al. (2009) simulation results, we find that for every micrometeorite that hits Vesta, about 260 hit the Moon. Scaling from the lunar flux, these rates imply that primitive micrometeorites add an amount of DM to Vesta of $2\times10^{-12}$ meters per year. Thus, even over a timescale of a few Gy, the depth of DM added to Vesta is less than a cm, not enough to explain the observed dark layer deposits. In addition, this material is added incrementally, giving impact processes an opportunity to mix it into Vesta's regolith. This mixing would further dilute any layer produced, though it could generally lower Vesta's albedo.

A potentially larger source of micrometeorites would be those comets that disrupted early in solar system history. Here we refer to the planetary evolution scenario known as the Nice Model which suggests that the migration of the giant planets about 4 Gy ago scattered and decimated a primordial disk of comets that was residing just outside the original orbits of Uranus and Neptune (Tsiganis et al. 2005; Gomes et al. 2005). This disk may have contained as much as 35 Earth masses of material prior to giant planet migration, much more than the 0.1 Earth masses of material found in the present-day trans-Neptunian scattered disk. Moreover, the interval when these comets were crossing the inner solar system was limited to a mere few tens of My, short enough that micrometeorites coming from this source might build a substantial dark layer of material on Vesta.

Assuming that scattered primordial comets disrupt in the same manner as JFCs do today, Nesvorny et al. (2010) predict that the mass of micrometeorite material added to the Moon during the Late Heavy Bombardment (LHB) in the form of CI/CM material was $2\times10^{20}$ g. This implies the amount of DM added to Vesta was $2\times10^{20}$ g / 260, or $7.6 \times 10^{17}$ g. If we assume micrometeorites have a bulk density of 1 g/cm$^3$, and that this material was added uniformly over Vesta's surface, we get a layer that is ~1 m thick. Impacts would gradually mix in and dilute this layer, at least until everything was buried by ejecta from some sufficiently large impact event.

A ~1 meter thick layer of dark material could potentially lead to observable effects on Vesta. The problem, however, is that it is unclear whether it is the right order of magnitude to match observations once regolith mixing and dilution had been carried out. It is also unclear whether observations are consistent with a global layer of DM hidden from sight by brighter layers. Thus, as before, we believe these values are interesting enough to warrant further work, but we lack definitive answers at this point. A related argument against a micrometeorite source is that a micrometeorite layer deposited over the course of solar system history would be uniformly distributed while the DM distribution on Vesta is non-uniform with localized concentrations. Meteoritical evidence also points to two distinct types of carbonaceous chondrite inclusions (macro clasts and micrometeorites) with two different mechanisms for their origin (Zolensky et al. 1996; Gounelle et al. 2003).

We conclude this section by pointing out that asteroid disruption events during the earliest times of main belt evolution could provide an additional source of primitive micrometeorites to Vesta (e.g., Bottke et al. 2005; Levison et al. 2009; Rubin and Bottke 2009) apart from impact of dark asteroids. The strength of the source has yet to be assessed in a quantitative fashion. The hope is that Vesta constraints can be used to rule this source in or out.

## 7. Conclusion

Dawn FC color images show that DM is ubiquitous on Vesta and preferentially enriched in and around the rim of the ancient Veneneia basin. The composition and distribution of DM on Vesta is consistent with Veneneia formation by a low-V carbonaceous impactor. In respect to geomorphology, the distribution of DM seems to be mainly linked to impact craters with DM found around the crater rims in ejecta blankets, forming ejecta rays, on crater rims, in debris flows on crater walls and sometimes deposited on crater floors, and likely in small secondary impact craters or crater chains. In addition, Aricia Tholus is an impact crater with dark ejecta rays that looks peculiar because it is placed on a hill, but the nature of Lucaria Tholus with its dark flow features remains to be confirmed.

Comparison of color parameters of various meteoritical analogs suggests that carbonaceous clast-rich howardites (Buchanan et al. 1993; Zolensky et al. 1996) such as PRA04401 (Herrin et al. 2011) are the best analogs for the DM. This result also confirms ground-based telescopic observations of Vesta that suggested contamination from impacting carbonaceous chondrites (Hasegawa et al. 2003; Rivkin et al. 2006). Our analysis of endogenous sources of DM, such as pyroclastic deposits (Wilson and Keil, 1996), impact shock/melt (Burbine et al. 2001), and eucrites rich in opaque minerals (Mayne et al. 2010) indicates that these are unlikely sources of DM. Petrologic models (Wilson and Keil, 1996) and meteorite studies (Mayne et al. 2010) also support our results. The abundance of CM2 carbonaceous chondrite material on the surface of Vesta estimated by Dawn FC data (<6 vol.%) is consistent with HED meteorites (Zolensky et al. 1996; Herrin et al. 2011). This value is also consistent with estimates of CM-like material (<10%) proposed by Rivkin et al. (2006) to explain the 3-µm absorption feature. The connection between Vesta, Vestoids and HED meteorites is primarily based on near-IR spectral observations and dynamical modeling of the delivery mechanism (Binzel and Xu, 1993). The association between dark material on Vesta and carbonaceous chondrite clasts in howardites provides the first direct and independent evidence that these meteorites were actually delivered from Vesta.

**Acknowledgement**

We thank the Dawn team for the development, cruise, orbital insertion, and operations of the Dawn spacecraft at Vesta. The Framing Camera project is financially supported by the Max Planck Society and the German Space Agency, DLR. We also thank the Dawn at Vesta Participating Scientist Program (NNX10AR22G) for funding the research. Dawn data is archived with the NASA Planetary Data System. EAC thanks CFI, MRIF, CSA, and the University of Winnipeg for support of this project and the necessary infrastructure. We thank Paul Buchanan (Kilgore College) and Tom Burbine (Mt Holyoke College) for their helpful reviews to improve the manuscript. VR would like to thank Richard Binzel (MIT), Rhiannon Mayne (TCU), and Guneshwar Thangjam (MPS) for their helpful suggestions to improve the manuscript.

**Figure Captions**

**Figure 1.** (A) Mosaic of 0.750 μm filter HAMO images (60 m/pixel) of the 58-km diameter Marcia crater showing downslope movement of bright material covering the underlying DM in periodic mass-wasting events probably trigged by seismic shaking. In the close-up view showing a LAMO clear filter image at 20 m/pixel (B), discontinuous clumps of material along the crater wall seem to channel the bright material (dashed white arrows) around the DM deposits, creating 'tadpole' like DM features (white arrows). North is up in all images. All coordinates are based on the new Claudia

coordinate system, which is different from the older Olbers system used with the HST data. Olbers reference longitude is located approximately at 210°E in the Claudia system.

**Figure 2.** (A) Mosaic of 0.75-µm filter HAMO images (60 m/pixel) of the Numisia impact crater (7°S, 247°E) showing downslope movement of DM. The northern dark debris flow (B) is likely similar to the ejecta blanket seen to the north of the crater rim. Like for Marcia crater, discontinuous clumps of bright material along the crater wall are channeling the DM around them. (B) is a LAMO clear filter image at 20 m/pixel. (C) Close-up view (0.75-µm filter LAMO image) of the northwest part of the rim and crater wall of Numisia showing bright outcrops, dark deposits (white arrows) and bright to medium albedo debris flows (dashed white arrows). North is up in all images.

**Figure 3.** Impact craters Occia (A) and Vibidia (B) showing mixture of DM with bright material. (C) Small impact crater (14.9°N, 179.9°E) on the ejecta blanket of the impact crater Marcia showing DM ejecta radially distributed. (D) Dark spots associated with two shallow impact craters (the one on the right is located at 27.8°S, 152.6°E). Some of these dark spots are likely secondary impact craters. Image (A) was obtained during the LAMO phase at a resolution of 20 m/pixel and images (B-D) during the HAMO phase at a resolution of 60 m/pixel. (A-D) are photometrically corrected 0.75 µm-filter images. North is up in all images.

**Figure 4.** (A) Radial ejecta of DM from a crater on Aricia Tholus (12.1°N, 161.8°E), a 37-km diameter hill shown overlaid on topographic map in Fig. 4 (B) and (C). These clear filter images were obtained during the HAMO phase at a resolution of 60 m/pixel.

**Figure 5.** (A) Lucaria Tholus, a 22-km diameter ridge (13°S, 104°E) showing DM deposits along the ridgeline. Figures 5 (B-D) show the image wrapped on a topographic map to give vertical perspective of the feature. These clear filter images were obtained during the HAMO phase at a resolution of 60 m/pixel.

**Figure 6.** (A) Image at 0.75 μm of two impact craters showing DM from RC3 phase at a resolution of 487 m/pixel. The crater near the top of the image is named Rubria and the one near the bottom is Occia. (B) Profile of the albedo from the red line drawn in (A) on the Occia crater. A distinct drop in albedo is seen across the DM ejecta.

**Figure 7.** (a) 0.75-μm filter image of 15-km diameter impact crater Cornelia (9.3°S, 225.5°E) showing the DM mixed with bright materials in the crater walls. (b) Band depth color-coded image *(R(0.75 μm)/R(0.92 μm))* of the same crater showing weaker band depth for DM (blue) and deeper band depth for fresh bright material (red). Albedo and band depth are the only two color parameters affected by the presence of DM on the surface. (C) Eucrite-Diogenite ratio (0.98/0.92 μm) image showing areas rich in diogenite (red) and those more eucrite-like (deep blue). Howardite-rich areas appear more yellow-cyan. The ED ratio is calculated based on *R(0.98)/(0.92)*, where R is the reflectance at wavelength λ.

**Figure 8.** Band depth vs. albedo plot with the vestan DM and three opaque-rich eucrites and quenched eucrite identified by Mayne et al. (2010). The band depth and albedo of all four eucrites are higher than those of the DM found in our Cornelia crater test site. Error bars plotted for the DM are one standard deviation.

**Figure 9.** Band depth vs. albedo plot with the vestan DM and shocked eucrite JaH 626 in three grain size ranges, shocked eucrite Padvarninkai in two grain size ranges, eucrite Macibini impact melt clast and impact melt matrix from polymict eucrite LEW85303. The band depth and albedo of all samples differ from those of the DM found in our Cornelia crater test site.

**Figure 10.** (a) Comparison of color spectra of DM (black) and laboratory intimate mixtures of <45 μm size CM2 carbonaceous chondrite Murchison and eucrite Millbillillie normalized to unity at 0.75-μm. Average DM spectrum plots between 10-20 wt.% CM2 abundance spectra. (b) Albedo vs. band depth plot showing DM from Cornelia crater (open circles), areal (solid gray line) and intimate (dotted gray line) mixtures of CM2+eucrite, and carbonaceous chondrite-rich howardite PRA04401 (yellow circle). The dark material trend falls in between areal and intimate mixtures of CM2+eucrite suggesting that both areal and intimate mixing has occurred on the surface of Vesta. The match between PRA04401 and DM along with similar albedo vs. band depth trends with CM2+eucrite mixtures suggests that at least some DM on Vesta is due to infall of carbonaceous chondrite meteorites.

**Figure 11.** Distribution of geologic features associated with DM plotted on a 0.75 μm filter mosaic of Vesta obtained at a resolution of ~480 and ~60 m/pixel. Dark material is divided in three categories with spots, craters and topographic highs. Craters marked with the circle symbol exhibit DM either on the crater wall in the form of linear debris flows, avalanches or V-shaped deposits, on crater rim or in the ejecta blanket (ejecta rays and/or units draping the surrounding terrain). Some craters have DM both inside the crater and outside in the ejecta. Many of the dark spots (white diamond) are likely secondary impact craters, even though not all can be traced back to the primary impact, and that are so small that the detection of their crater rim is uncertain. The last type is represented here by the only topographic high visible in the southern hemisphere, Lucaria Tholus (star). Along with those features, the approximate outline of the rim of the Rheasilvia basin (black line) and the incomplete rim of the Veneneia basin (red line) are depicted on the map. Two large dark albedo areas forming a regional mantling and corresponding to the ejecta of Marcia (10°N, 190°E) and Octavia (3.3°S, 147°E) craters are not specifically mapped but are included in the crater category (represented by a circle for each crater).

**Figure 12.** Global distribution of geologic features associated with DM displayed onto a color-coded and hill shaded topographic map of Vesta in cylindrical projection. Elevation is computed relative to a 285 x 229 km reference ellipsoid with blue corresponding to the lowest elevation and red being the highest. Dark material is divided in three categories with spots (white diamond), craters (circle) and topographic highs (star). Along with those features, the approximate outline of the rim of the Rheasilvia basin (black line) and the incomplete rim of the Veneneia basin (red line) are depicted on the map. Data was trimmed in illumination angles (diagonal lines area) so DM was not mapped at the extreme South Pole.

**Figure 13.** (a) Collisional velocity distributions for Vesta and the entire main belt, calculated using the Bottke et al. (1994) algorithm and the orbital distribution of all asteroids larger than 30 km diameter. The velocity distribution of asteroidal impacts on Vesta has a mean impact velocity of 4.75 km/s (RMS 5.18 km/s, median 4.41 km/s), which is comparable to but somewhat lower than the calculated mail-belt average value of 5.14 km/s (O'Brien and Sykes 2011). While the distribution has a mean value, impacts can occur at significantly higher and lower values. For reference, Vesta's escape velocity is approximately 350 m/s. (b) Same as (a) but integrated to show the probability of impacts occurring at less than a given velocity for Vesta and the entire main belt. Figures are adopted from O'Brien and Sykes (2011).

**Figure 14.** Carbonaceous chondrite volume fraction in the crater ejecta for a range of target materials, impactor diameters and velocities, calculated using the methods described in Sec. 6.1.2. Note that the scaling laws for hard rock and soft rock overlap for impactor diameters larger than a few km, as do the laws for dry soil and sand.

**Figure 15.** Results of a dynamical model of the ejecta distribution from the Veneneia basin on Vesta using methods described in Sec. 6.1.3 and Durda et al. (2004).

**Table 1. Color parameters of dark material along with possible analogs**

| Analogs/Effects | Albedo | Band Depth | Examples |
|---|---|---|---|
| Dark Material | Low | Weak | Vesta |
| Volcanic (pyroclasts) | Low-Moderate | Weak | Pyroclastic deposits (Moon) |
| Shock/Impact Melt | Moderate | Moderate | Moon/Vesta |
| Eucrites with Opaques | Moderate | Moderate | Chervony Kut meteorite |
| Exogenous Material (carbonaceous chondrites) | Low | Weak | CM2, CR clasts in HEDs |

**Table 2. Percentages of total impacts occurring below a given velocity (km/sec). For reference, Vesta's escape velocity is approximately 350 m/sec.**

| Velocity (km/sec) | Fraction of impacts less than or equal to V (%) |
|---|---|
| 1 | 0.42 |
| 2 | 5.8 |
| 3 | 19.6 |
| 4 | 40.8 |
| 5 | 60.2 |

**Figure 1: Dawn at Vesta, Dark Material**

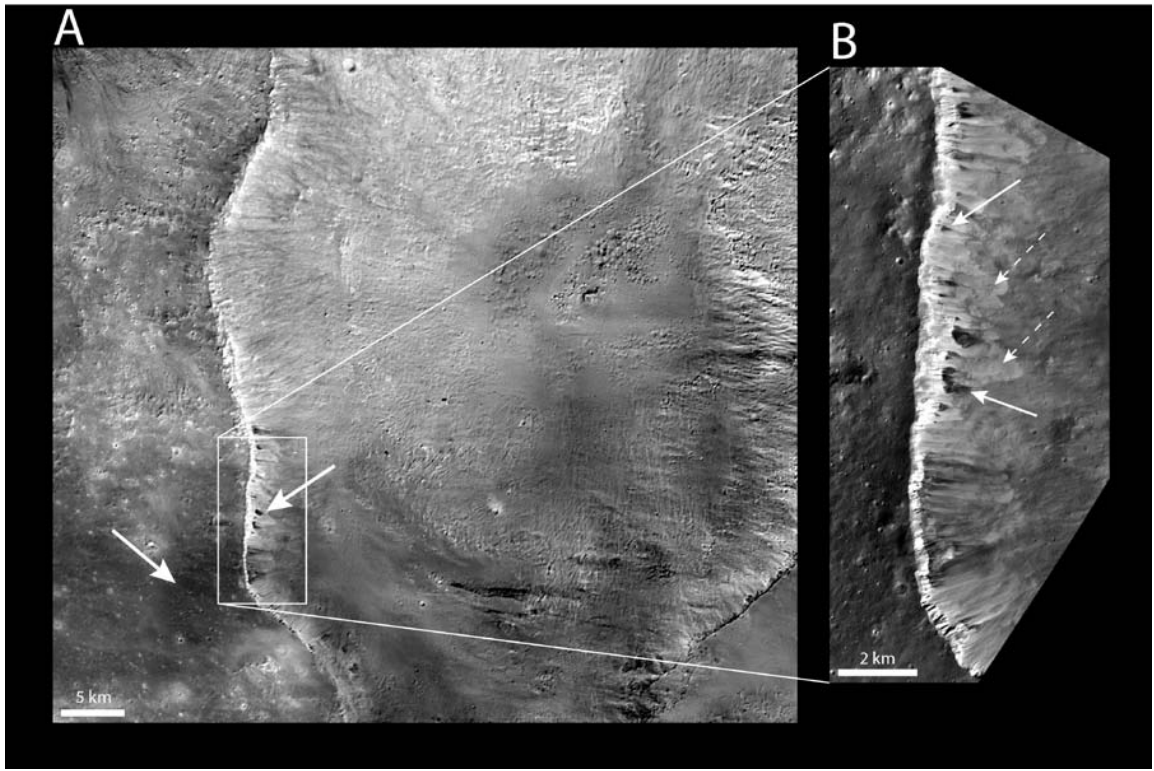

**Figure 2: Dawn at Vesta, Dark Material**

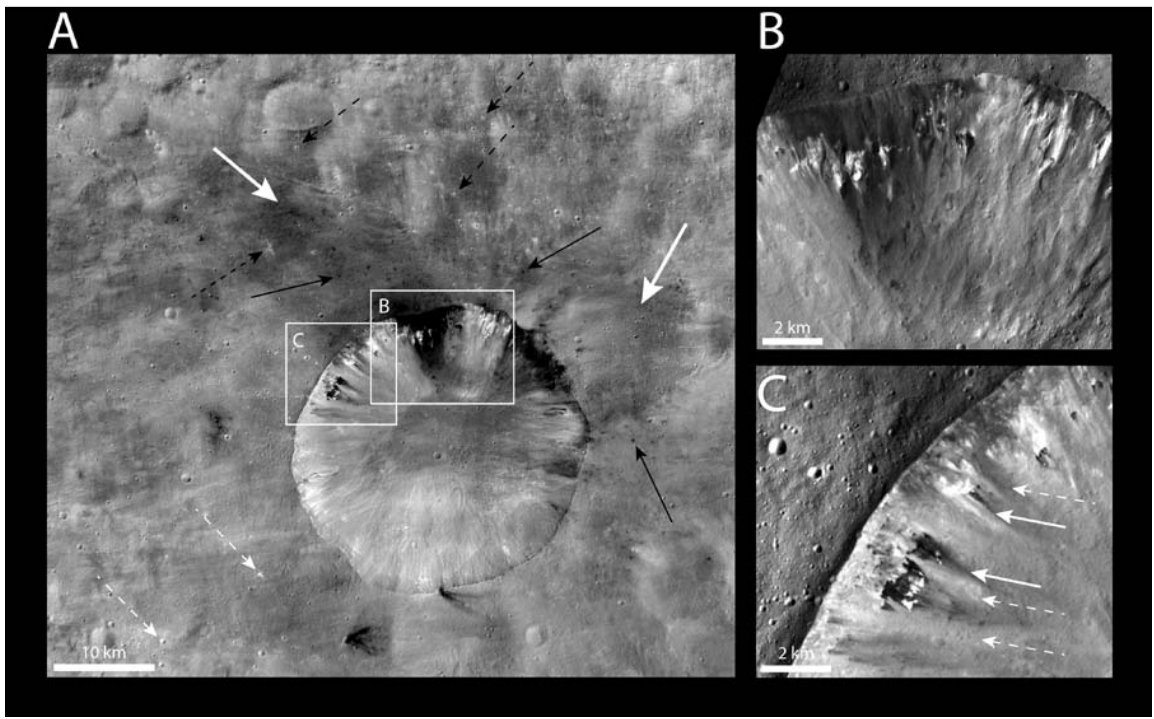

**Figure 3: Dawn at Vesta, Dark Material**

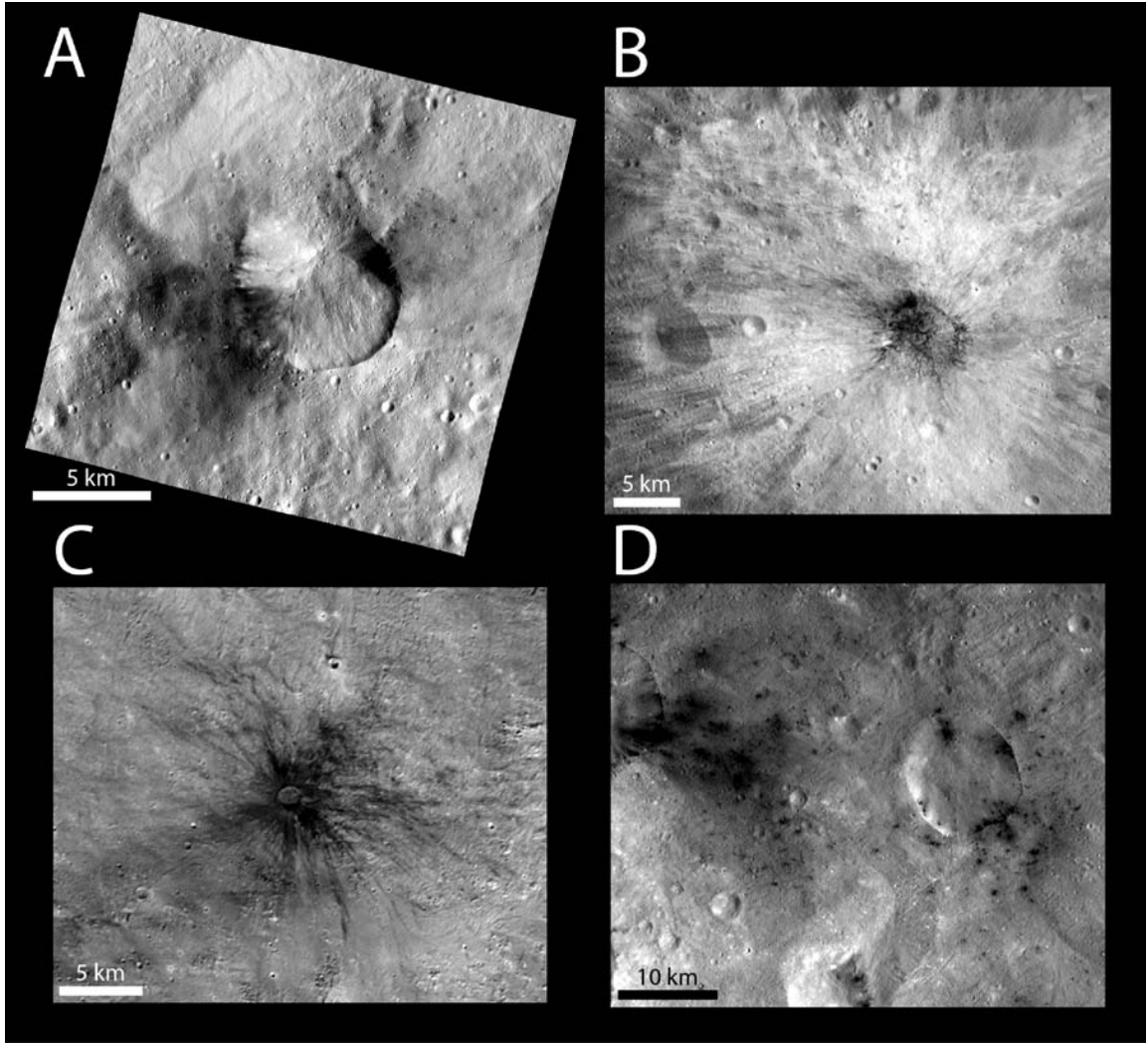

**Figure 4: Dawn at Vesta, Dark Material**

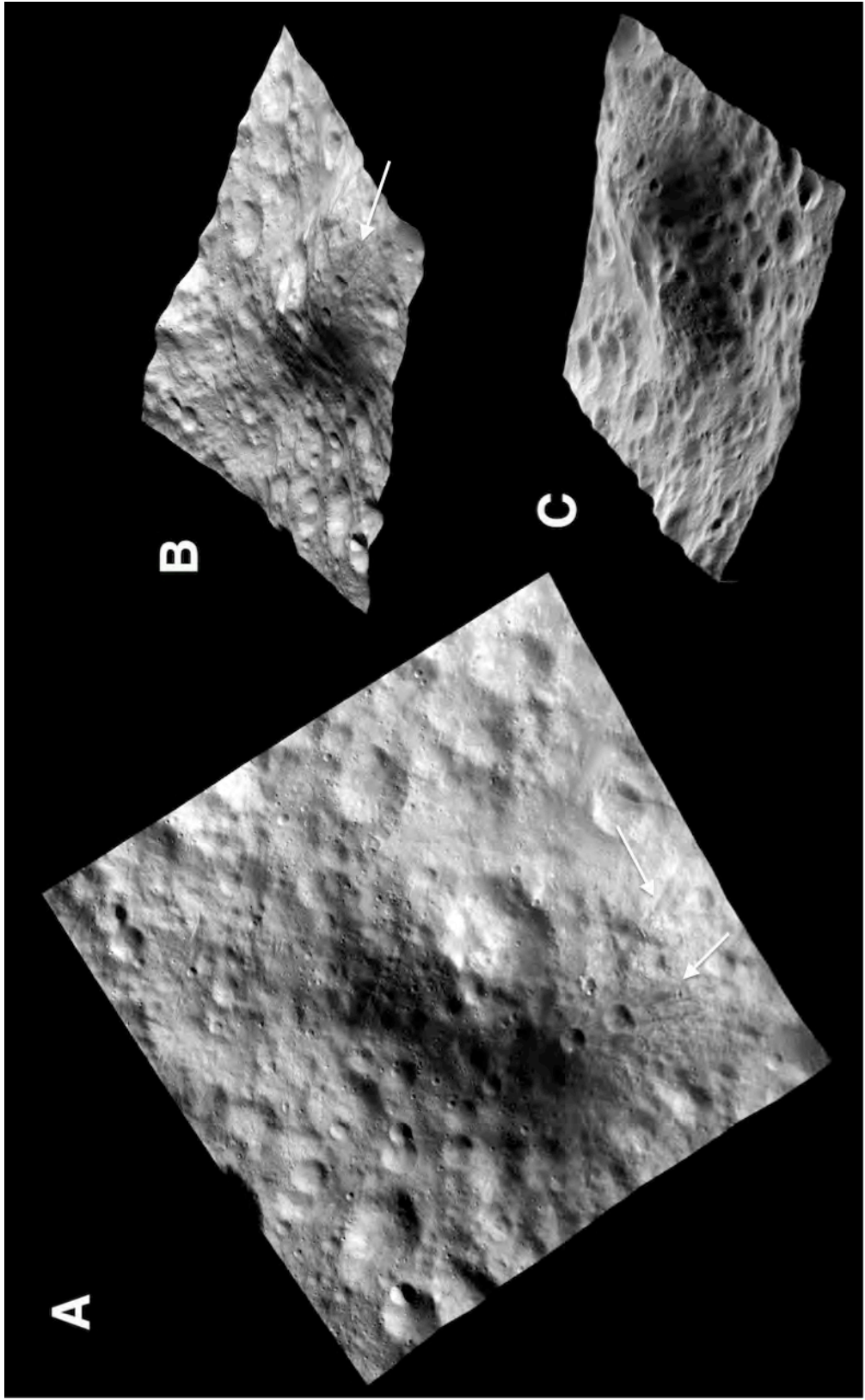

**Figure 5: Dawn at Vesta, Dark Material**

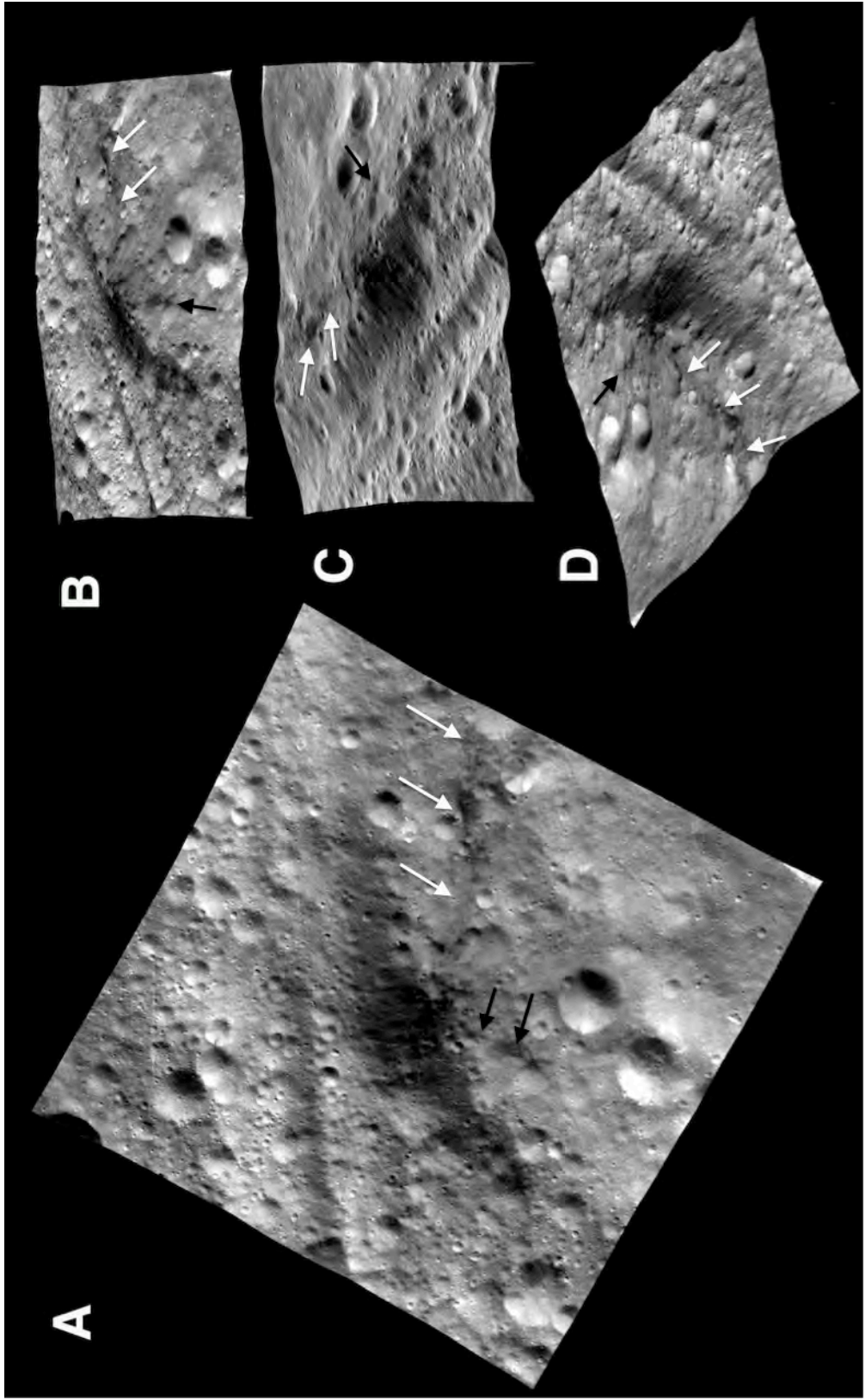

**Figure 6: Dawn at Vesta, Dark Material**

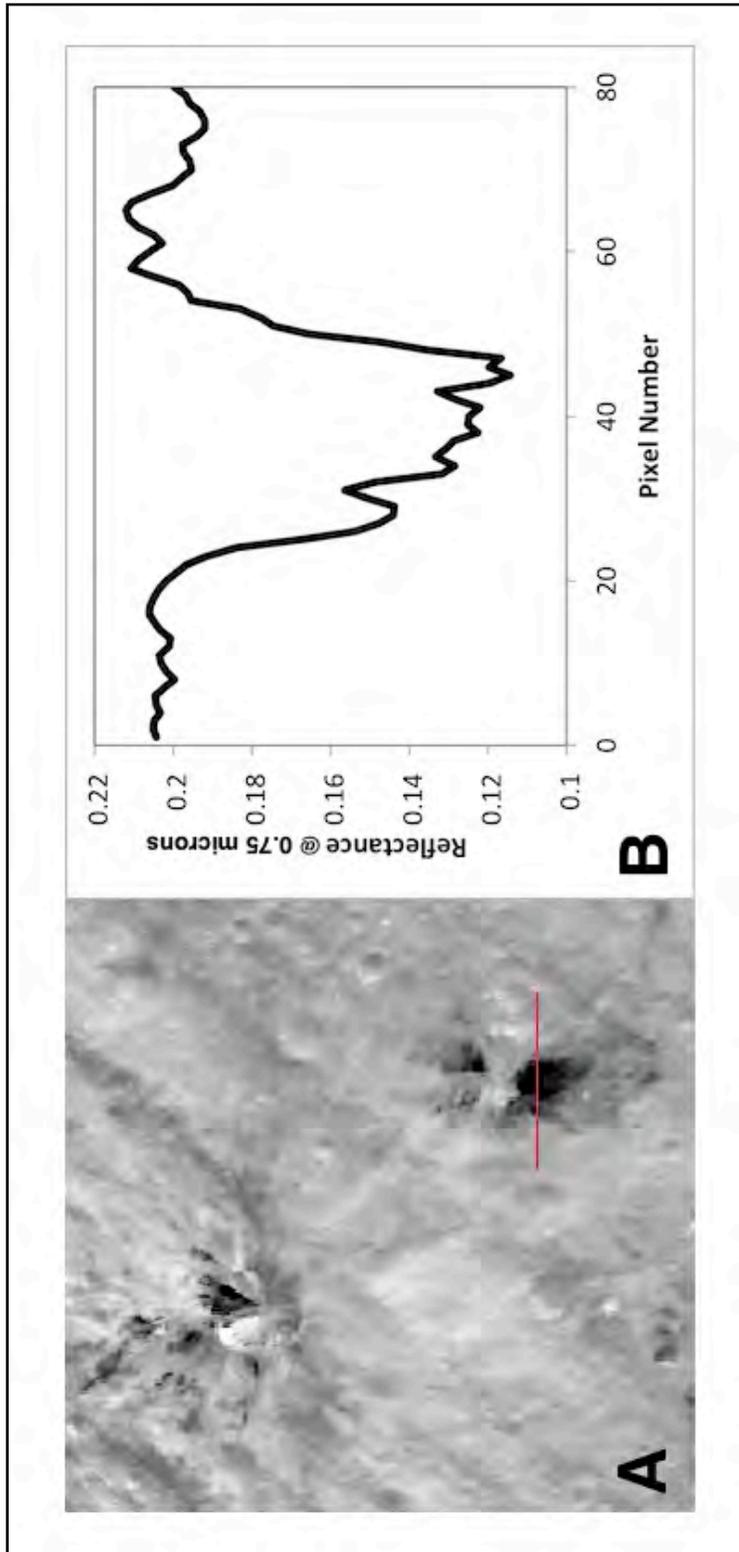

**Figure 7: Dawn at Vesta, Dark Material**

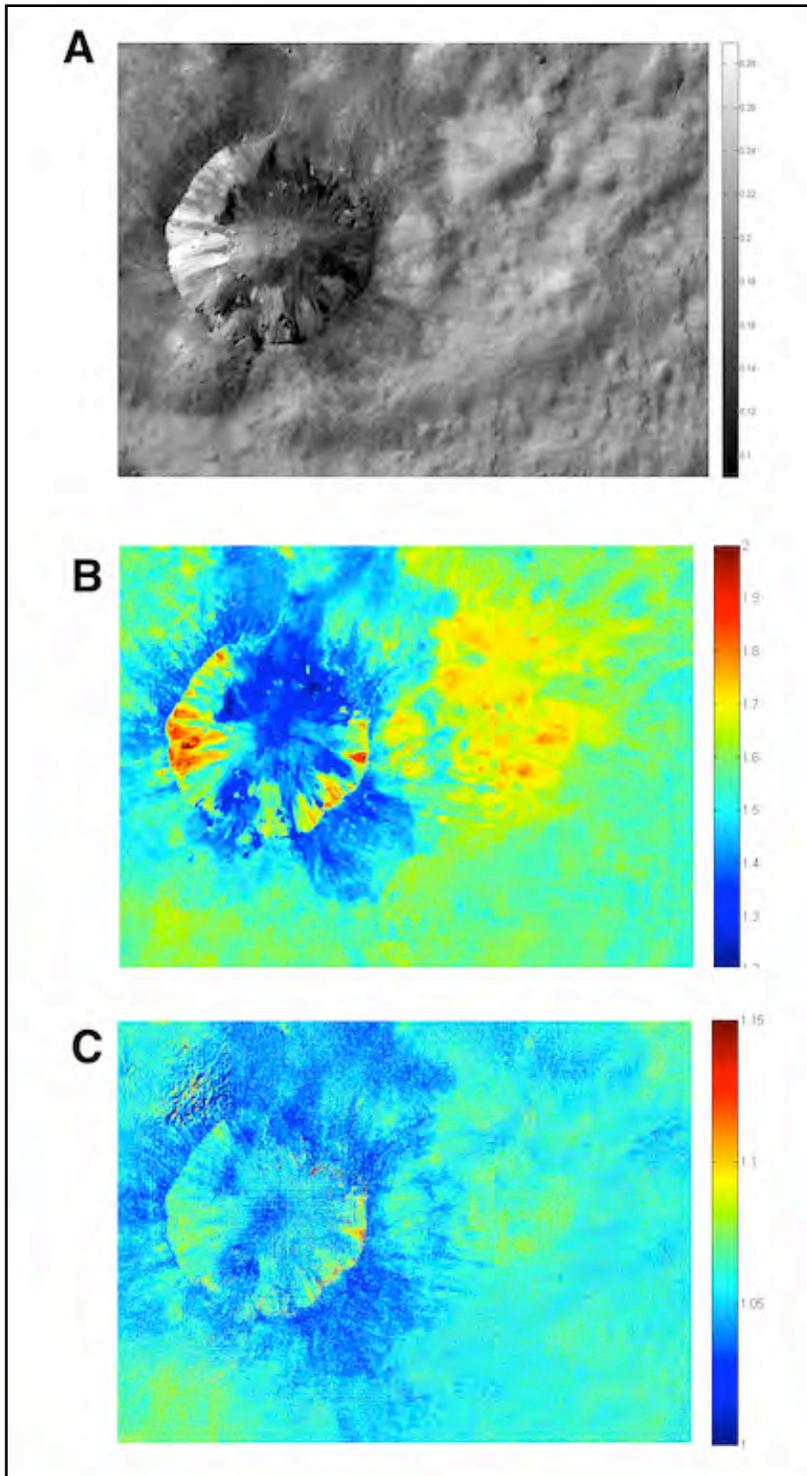

**Figure 8: Dawn at Vesta, Dark Material**

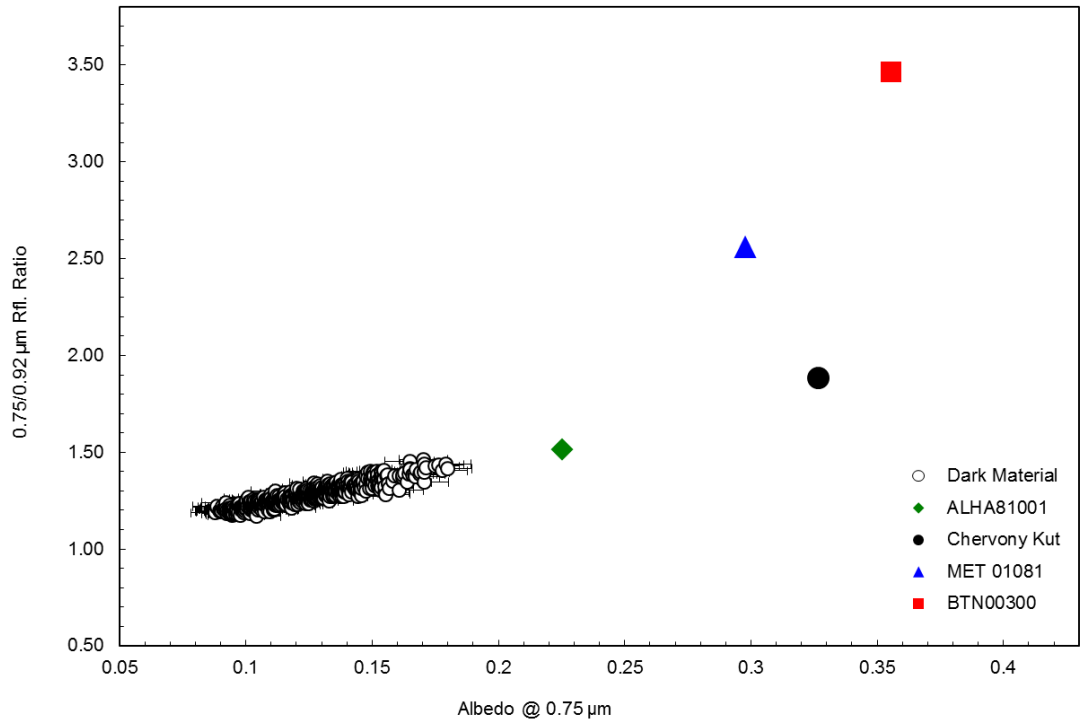

**Figure 9: Dawn at Vesta, Dark Material**

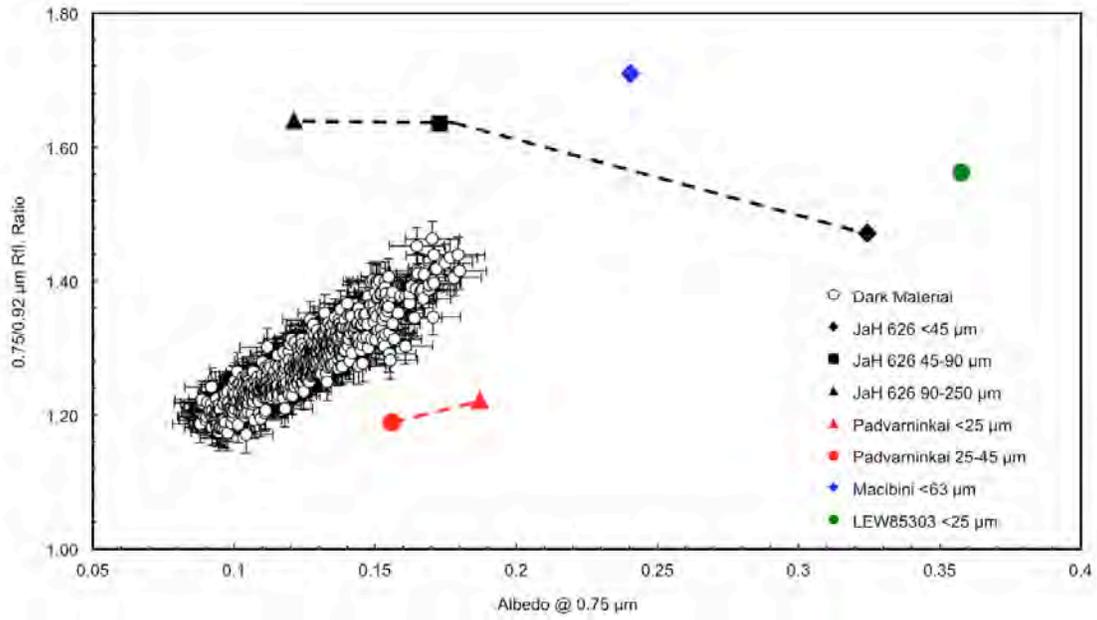

**Figure 10: Dawn at Vesta, Dark Material**

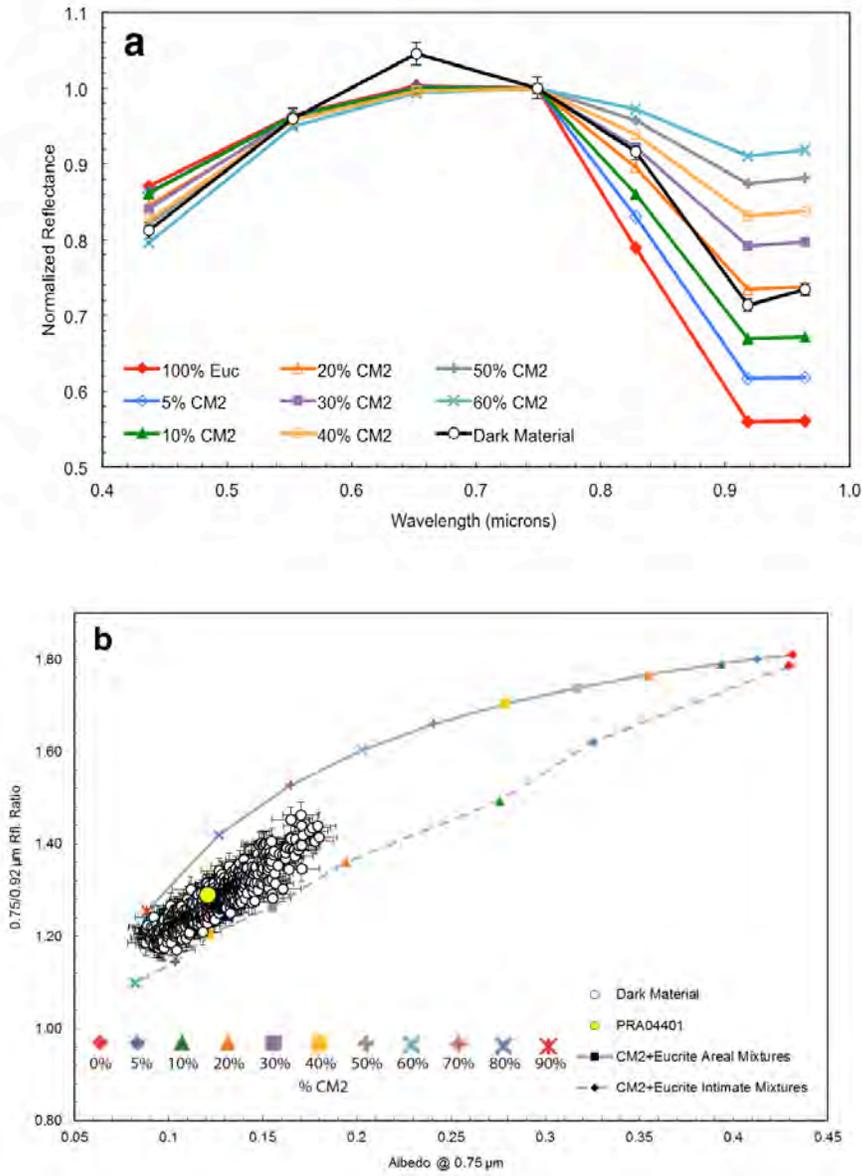

**Figure 11: Dawn at Vesta, Dark Material**

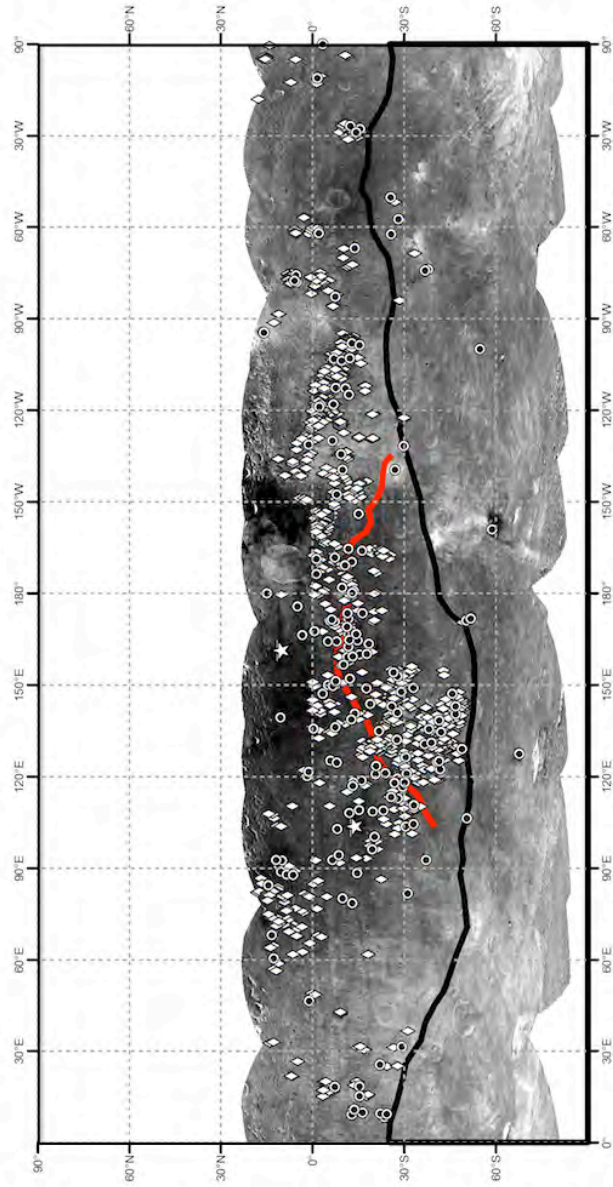

**Figure 12: Dawn at Vesta, Dark Material**

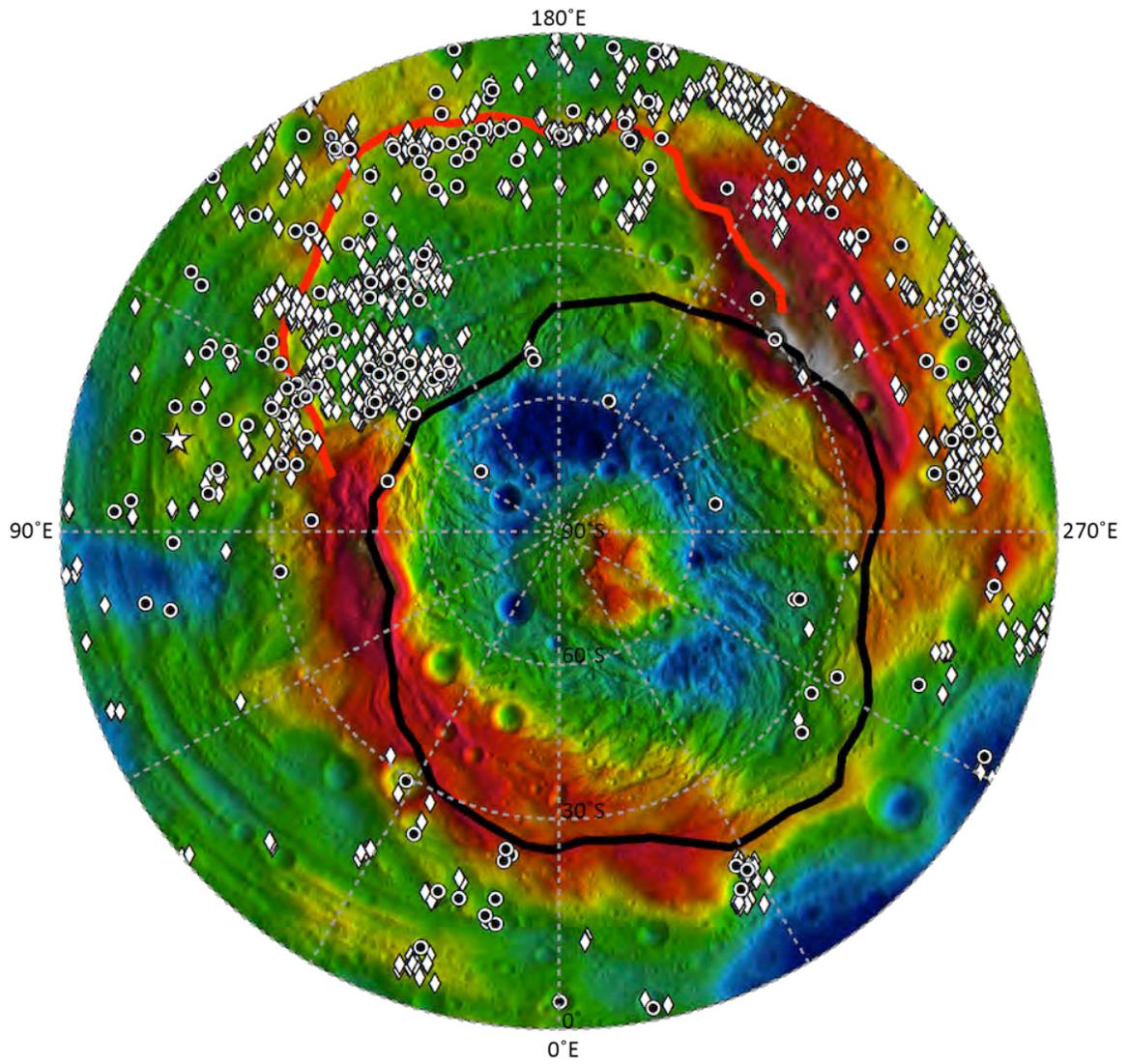

**Figure 13: Dawn at Vesta, Dark Material**

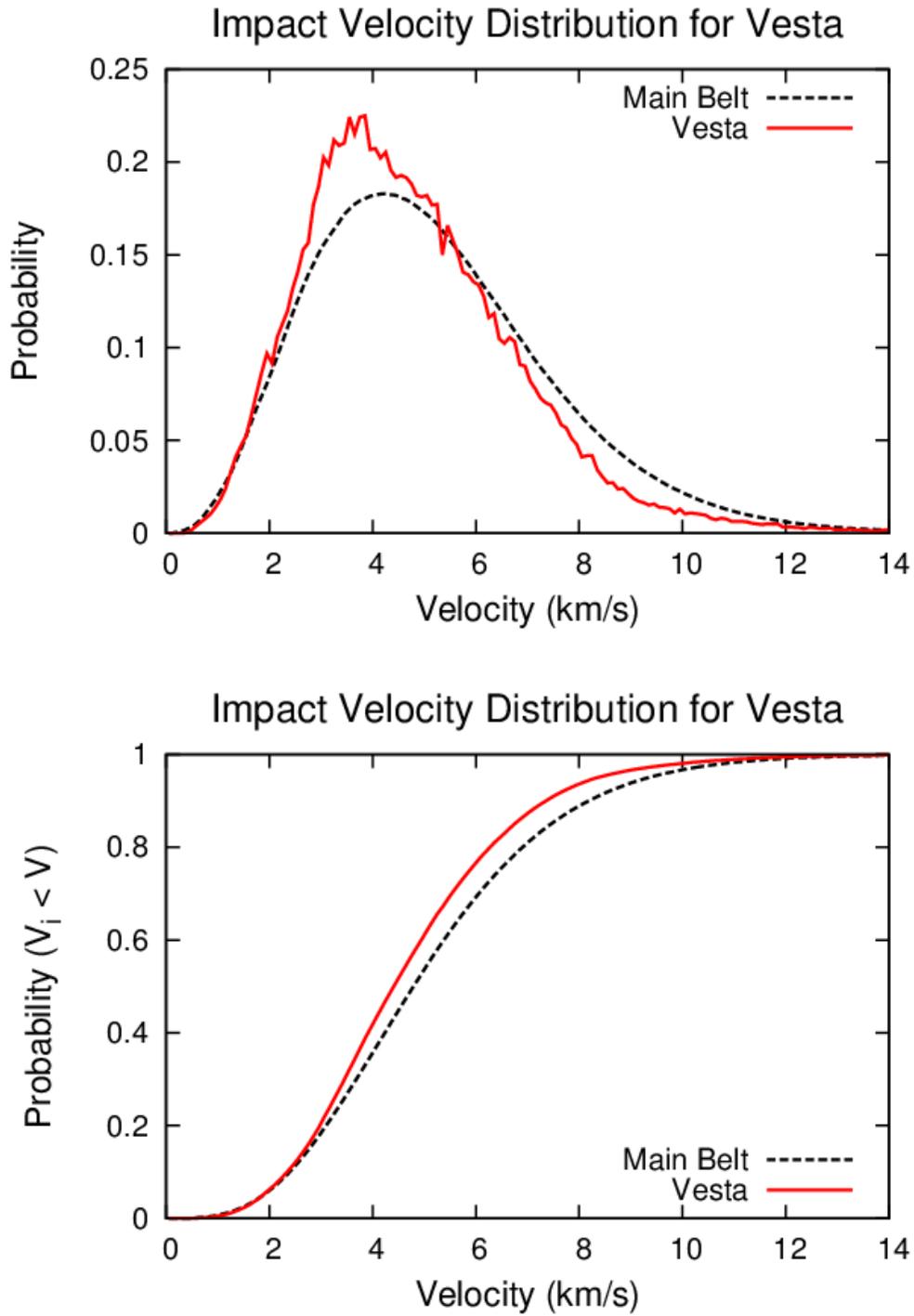

**Figure 14: Dawn at Vesta, Dark Material**

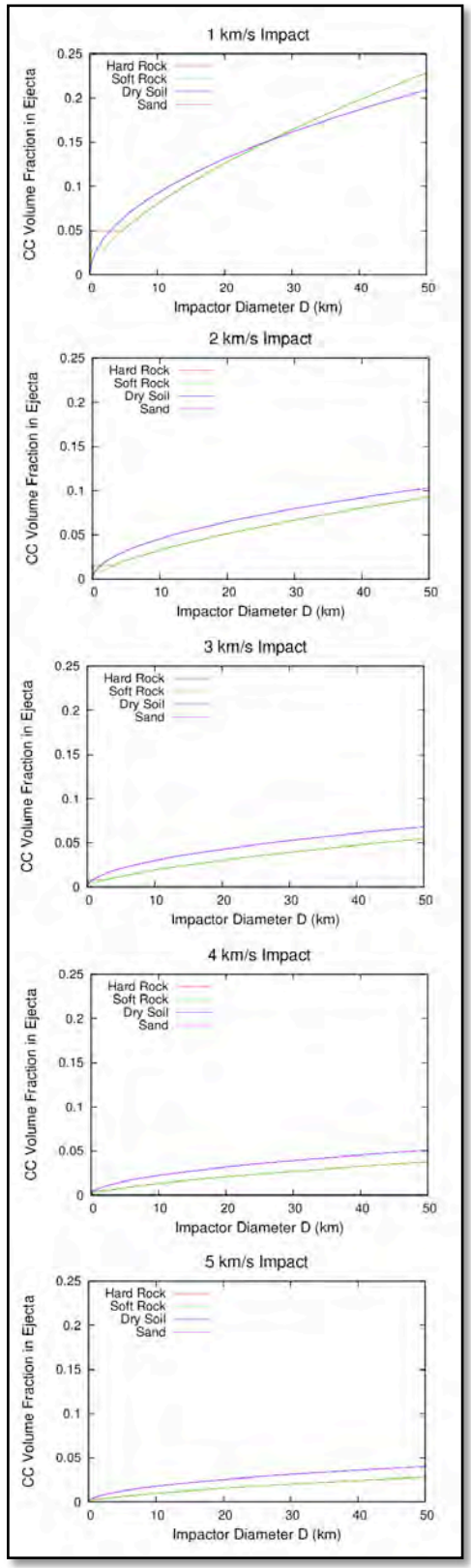

**Figure 15: Dawn at Vesta, Dark Material**

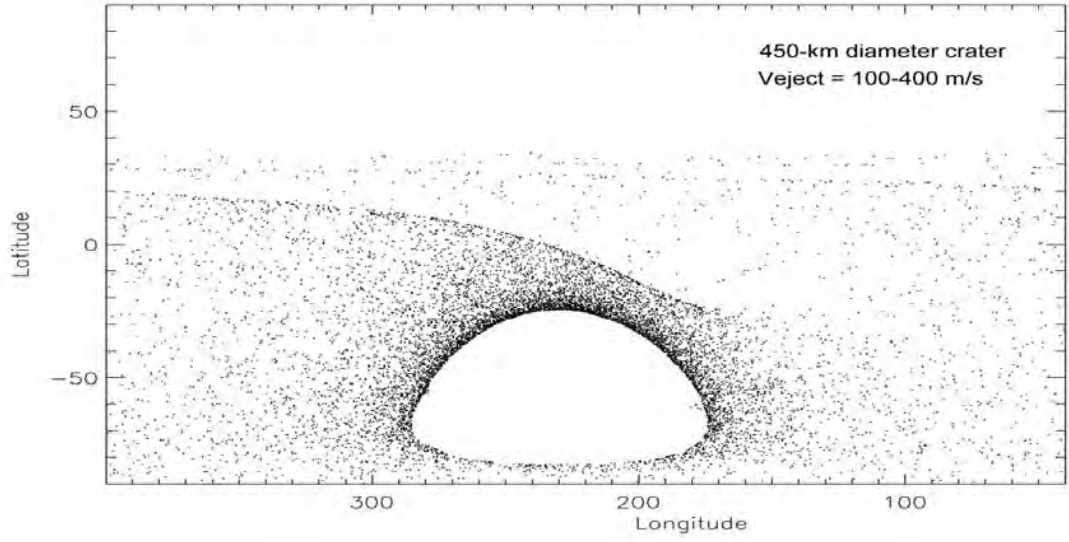